\documentclass[
 reprint,
 superscriptaddress,
 amsmath,amssymb,
 aps,
 prc,
 floatfix,
]{revtex4-2}

\usepackage{graphicx}
\usepackage{dcolumn}
\usepackage{bm}
\usepackage{xspace} 
\usepackage[utf8]{inputenc}

\begin{document}

\newcommand{\roots}    {\ensuremath{\sqrt{s}}}
\newcommand{\jT}         {\ensuremath{j_{\mathrm{T}}}\xspace}
\newcommand{\pT}         {\ensuremath{p_{\mathrm{T}}}\xspace}
\newcommand{\detastar} {\ensuremath{\Delta\eta^{\ast}}\xspace}
\newcommand{\dphistar} {\ensuremath{\Delta\phi^{\ast}}\xspace}
\newcommand{\etastar} {\ensuremath{\eta^{\ast}}\xspace}
\newcommand{\phistar} {\ensuremath{\phi^{\ast}}\xspace}
\newcommand{\vstarTWO} {\ensuremath{v^{\ast}_2}\xspace}
\newcommand{\VstarNdelta} {\ensuremath{V^{\ast}_{\mathrm{n}\Delta}}\xspace}
\newcommand{\VstarONEdelta} {\ensuremath{V^{\ast}_{\text{1}\Delta}}\xspace}
\newcommand{\VstarTWOdelta} {\ensuremath{V^{\ast}_{\text{2}\Delta}}\xspace}
\newcommand{\VstarTHRdelta} {\ensuremath{V^{\ast}_{\text{3}\Delta}}\xspace}
\newcommand{\AonA}  {\ensuremath{\text{AA}}\xspace}
\newcommand{\nchj}{\ensuremath{N_\text{ch}^\mathrm{j}}}
\newcommand{\nchtrg}{\ensuremath{N_{\text{ch}}^{\text{trg}}}}
\newcommand{\bdist}{\ensuremath{B(\detastar,\dphistar)}}
\newcommand{\sdist}{\ensuremath{S(\detastar,\dphistar)}}
\newcommand{\zgrg}{\ensuremath{z_g\theta_g^\beta}}

\title{On the Origin of QCD Collectivity in High-Multiplicity Jets:\\ A Transport Model Study}

\author{Xiao Huang}
\affiliation{Department of Physics and Astronomy, Rice University, Houston, Texas 77005, USA}
\author{Xiaoyu Liu}
\affiliation{Department of Physics and Astronomy, Rice University, Houston, Texas 77005, USA}
\author{David Qian}
\affiliation{Saint John's School, Houston, Texas 77019, USA}
\author{Wei Li}
\affiliation{Department of Physics and Astronomy, Rice University, Houston, Texas 77005, USA}
\author{Zi-Wei Lin}
\affiliation{Department of Physics, East Carolina University, Greenville, North Carolina 27858, USA}
\author{Xin-Nian Wang}
\affiliation{Key Laboratory of Quark and Lepton Physics (MOE) and Institute of Particle Physics,
Central China Normal University, Wuhan 430079, China}
\author{Wenbin Zhao}
\affiliation{Key Laboratory of Quark and Lepton Physics (MOE) and Institute of Particle Physics,
Central China Normal University, Wuhan 430079, China}

\date{\today}

\begin{abstract}
The CMS Collaboration has observed an enhancement of elliptic azimuthal anisotropy (\vstarTWO) in high-multiplicity jets. To investigate its microscopic origin, we employ a hybrid transport model that couples jets generated with \textsc{PYTHIA~8} to partonic and hadronic rescattering. The analysis is performed in the jet frame, where the jet momentum defines the longitudinal axis. We characterize the initial-state geometry using the eccentricity vectors of shower partons and quantify the geometric response by correlating them with the final-state flow vectors of hadrons. By systematically varying the partonic and hadronic interactions, we find that the anisotropy enhancement is dominated by hadronic rescattering in the present model and increases with the initial eccentricity. We further classify jets using the Soft Drop variable \zgrg and show that this momentum-space substructure variable is sensitive to the initial coordinate-space geometry, although the predicted substructure dependence differs from the current CMS measurement. These results support a geometry--response mechanism for collective behavior inside jets and establish jet substructure as a promising experimental handle on the initial geometry. They also motivate models that treat parton branching and transport interactions concurrently.
\end{abstract}

\maketitle


\section{\label{sec:intro}Introduction}
Collective phenomena are among the most remarkable emergent properties of strongly interacting QCD matter. In relativistic nucleus-nucleus ($\mathrm{AA}$) collisions, long-range near-side correlations, commonly known as the ridge, provide compelling evidence that the produced quark-gluon plasma (QGP) behaves collectively~\cite{Adams:2005ph,Alver:2008ck,Abelev:2009af,Alver:2009id,Chatrchyan:2011eka,Chatrchyan:2012_PbPbDihadron,Aamodt:2010pa,Aad:2012bu,Chatrchyan:2013nka}. A central paradigm established over the past two decades is that strong final-state interactions convert the initial spatial anisotropy of the collision geometry into final-state momentum anisotropy~\cite{Bjorken:1982qr,Alver:2010gr,Gardim:2011xv,Luzum:2013yya,Noronha:2024dtq}.

Surprisingly, similar ridge-like correlations have subsequently been observed in much smaller systems, including high-multiplicity proton-proton ($\mathrm{pp}$)~\cite{CMS:2010ifv,Aad:2015gqa,Khachatryan:2015lva,Khachatryan:2016txc,Aad:2019pfy} and proton-nucleus ($\mathrm{pA}$) collisions~\cite{Chatrchyan:2012wg,Abelev:2012ola,ATLAS:2012cix,Aaij:2015gna,Abelev:2013haa,Khachatryan:2014jra,Khachatryan:2015waa,Aaboud:2017acw,Aaboud:2017xfb,Aidala:2018mcw}. These discoveries suggest that collective behavior can emerge in systems containing only tens of final-state particles, raising two fundamental questions: What is the smallest QCD system capable of exhibiting collective behavior, and through what mechanisms does collectivity emerge from the fundamental interactions of QCD~\cite{Nagle:2013lja,Schenke:2014zha,Shen:2016zpp,Weller:2017tsr,Mantysaari:2017cni,Zhao:2017rgg,Bierlich:2019wld,Katz:2019qwv,Zhao:2020pty,Zhao:2022ayk,Zhao:2022ugy}?

Recent theoretical work~\cite{Baty:2021ugw} has explored these questions by postulating that collectivity may arise even from a single parton propagating in vacuum and fragmenting into a hadronic final state, either through strong rescattering reminiscent of a fluid system or through quantum entanglement. Motivated by this idea, the CMS Collaboration has reported the first evidence of collectivity inside high-multiplicity jets in proton-proton collisions: an enhancement of the long-range elliptic anisotropy \vstarTWO at high multiplicity~\cite{CMS:2023iam}. This observation opens a new window on the emergence of collective behavior in the smallest QCD system accessible experimentally.

Understanding the origin of this anisotropy naturally begins with the mechanism responsible for collective flow in heavy-ion collisions. In the conventional picture, strong final-state interactions convert the initial geometric eccentricity into the observed momentum anisotropy~\cite{Alver:2010gr,Gardim:2011xv,Luzum:2013yya}. Within the Bjorken picture~\cite{Bjorken:1982qr}, particles separated by large rapidity intervals remain correlated despite being causally disconnected during the later evolution because they inherit a common initial geometry. Whether the same mechanism operates inside a jet remains an open question.

In this work, we investigate this possibility using a hybrid transport model that combines PYTHIA 8-generated jets with partonic and hadronic rescatterings. By following the full space-time evolution of the jet shower, the model allows both the initial geometry and the subsequent final-state interactions to be studied within a common framework. This enables us to address three key questions:
\begin{itemize}
    \item What is the \textit{initial-state eccentricity} of a jet, and are
    eccentricities correlated across rapidity, as expected in a Bjorken-like picture?
    \item Can \textit{jet substructure} be used to engineer the initial eccentricity?
    \item Do the observed \textit{final-state anisotropies} arise from partonic
    or hadronic interactions, and do they retain correlations with the initial eccentricities?
\end{itemize}

By addressing these questions, this study aims to illuminate the microscopic 
origin of QCD collectivity inside high-multiplicity jets and provide insight 
into the formation of the smallest possible quark-gluon droplet in nature.

\section{\label{sec:model} Theoretical framework and model}
To investigate the emergence of collectivity inside high-multiplicity jets, 
we employ a hybrid transport model following the methodology of Ref.~\cite{Zhao:2024wqs}. 
This framework combines perturbative jet generation from \textsc{PYTHIA~8}~\cite{Sjostrand:2014zea} 
with partonic and hadronic transport evolution. The simulation proceeds in three stages~\cite{Lin:2004en}:
\begin{itemize}
    \item \textit{Initial-state generation and parton space-time structure:} 
    Jets are initially generated using \textsc{PYTHIA~8} (CP5 tune~\cite{Sirunyan:2019dfx}), 
    which provides both the hard scattering and subsequent parton shower. 
    Each shower parton is assigned a space--time formation point based on the full sequence of splittings from the hard scattering to the final-state parton.
    Specifically, the propagation time of each mother parton in the splitting sequence is estimated as~\cite{Adil:2006ra,Zhang:2022kuf} 
    \begin{equation}
    \label{eq:mother_propagation_time}
        \Delta t_i =
        \begin{cases}
            \dfrac{2E_i x_i(1-x_i)}{k_{T,i}^2}, & x_i < 0.5 \\[1ex]
            \dfrac{1}{E_i}, & x_i \ge 0.5
        \end{cases},
    \end{equation}
  where \(E_i\) is the energy of the mother parton in the $i$th splitting, while \(x_i\) and \(k_{T,i}\) are the fractional energy and transverse momentum of the daughter parton relative to the mother. The formation time of the final parton
  is the sum of the propagation times of all mother partons in the splitting sequence,
  after which the parton is allowed to interact with the medium of other shower partons. 
  The spatial coordinates are obtained as the sum of the displacements of all mother partons during their propagation times, with each mother parton assumed to propagate freely.

    \item \textit{Partonic rescatterings:}
    Once formed, partons undergo elastic two-body scatterings governed by the Boltzmann transport equation,
  \begin{equation}
  \label{eq:boltzmann}
      p^\mu \partial_\mu f(x,p) = C[f],
  \end{equation}
  where \(f(x,p)\) is the one-particle phase-space distribution and \(C[f]\) is the collision integral. The scattering kernel is implemented using Zhang's Parton Cascade (ZPC)~\cite{Zhang:1997ej,Lin:2001zk,Lin:2004en,Zhao:2020yvf}, with isotropic \(2 \to 2\) scattering and a constant partonic cross section \(\sigma_p\). The probability for two partons \(i\) and \(j\) to scatter in a time step \(\Delta t\) is
  \begin{equation}
  \label{eq:scatter_prob}
      P_{ij} = v_{\text{rel}} \, \sigma_p \, \frac{\Delta t}{V_{\text{cell}}},
  \end{equation}
  where \(v_{\text{rel}}\) is their relative velocity and \(V_{\text{cell}}\) is the local cell volume. Values of \(\sigma_p=0\)--$24~\mathrm{mb}$ are used to test the sensitivity to the rescattering strength.

    \item \textit{Hadronization and hadronic rescatterings:}
      After partonic rescattering ceases, the system is converted to hadrons using the Lund string-fragmentation model~\cite{Andersson:1983ia,Sjostrand:2007gs,JETSCAPE:2019udz,Zhao:2020wcd,Zhao:2021vmu,YuanyuanWang:2023vor}, as implemented in \textsc{PYTHIA}. A formation time of order \(\tau_{\text{had}} \simeq 1\,\mathrm{fm}/c\) is assumed for hadron production. The produced hadrons then undergo further evolution within the Ultrarelativistic Quantum Molecular Dynamics (UrQMD) model~\cite{Bleicher:1999xi}, which accounts for elastic and inelastic hadron-hadron scattering, resonance formation, and decay.
\end{itemize}
This hybrid approach allows us to compare results with and without final-state interactions. Pure \textsc{PYTHIA} showering captures only nonflow jet correlations, whereas rescattering can generate collective behavior. In particular, the model reproduces the increasing elliptic anisotropy \(v_2^*\) at high jet multiplicities (\(N_{\mathrm{ch}}^j \gtrsim 70\)). Within the present hybrid setup, the enhancement is dominated by hadronic final-state interactions, with partonic rescattering providing a smaller correction.

\section{Opacity, hydrodynamic response, and conformal scaling}
\label{sec:opacity_scaling}

We now derive the expected relation between opacity and final-state
multiplicity in a conformal system. The opacity is a dimensionless measure of
the interaction strength of an extended medium. It is defined as the
ratio of a characteristic path length to the microscopic mean free path,
\begin{equation}
  \hat{\gamma} \equiv \frac{L}{\lambda},
  \label{eq:opacity}
\end{equation}
where we identify the characteristic path length with the transverse system
size, \(L\sim R\).

For a weakly coupled conformal gas, temperature is the only dimensionful scale. Dimensional analysis then gives
\(n_0\propto T_0^3\) and, at fixed coupling, \(\sigma(T_0)\propto T_0^{-2}\). Consequently,
\begin{equation}
  \lambda_0 \sim \frac{1}{\sigma(T_0)n_0}\propto \frac{1}{T_0},
\end{equation}
where the subscript \(0\) denotes quantities evaluated at a reference initial
proper time \(\tau_0\). Eq.~\eqref{eq:opacity} then gives
\begin{equation}
  \hat{\gamma} \sim R T_{0}.
  \label{opacity_temperature}
\end{equation}

To express \(T_0\) in terms of an observable, we assume an effective 1D Bjorken longitudinal expansion during the early stage of jet evolution, where the initial entropy per unit rapidity is proportional to the final charged-particle multiplicity.
\begin{equation}
  \frac{dN_{\text{ch}}}{d\eta}
  \;\propto\;
  \frac{dS}{d\eta}
  \;\sim\;
  s_0 A_T \tau_0
  \;\sim\;
  T_0^3 A_T \tau_0,
\end{equation}
where \(s_0\propto T_0^3\) follows from conformal thermodynamics and
\(A_T\propto R^2\) is the transverse area. Solving for \(T_0\) gives
\begin{equation}
  T_0 \sim
  \left(
    \frac{1}{A_T\tau_0}
    \frac{dN_{\text{ch}}}{d\eta}
  \right)^{1/3}.
\end{equation}
Substituting this temperature relation into Eq.~\eqref{opacity_temperature}, the opacity becomes
\begin{equation}
  \hat{\gamma}
  \sim
  R\left(
    \frac{1}{A_T\tau_0}
    \frac{dN_{\text{ch}}}{d\eta}
  \right)^{1/3}
  \propto
  \left(\frac{dN_{\text{ch}}}{d\eta}\right)^{1/3}
  \left(\frac{R}{\tau_{0}}\right)^{1/3}.
  \label{eq:opacity_general_scaling}
\end{equation}

Thus, conformal microscopic and thermodynamic scaling alone leaves a residual
dependence on \(R/\tau_0\). In a scale-invariant or self-similar scenario where the initial formation time scales with the system radius ($\tau_{0}=c_{0}R$ with a constant $c_{0}$), the factor $(R/\tau_{0})^{1/3}=c_{0}^{-1/3}$ is constant. Under this condition, Eq.~\eqref{eq:opacity_general_scaling} reduces to
\begin{equation}
  \hat{\gamma} \sim
  \left(\frac{dN_{\text{ch}}}{d\eta}\right)^{1/3}.
  \label{eq:opacity_conformal_scaling}
\end{equation}
The multiplicity-only scaling in
Eq.~\eqref{eq:opacity_conformal_scaling} therefore follows from conformal
scaling together with self-similar initial conditions. If the response
coefficient \(v_2/\varepsilon_2\) is controlled primarily by the
opacity~\cite{Drescher:2007cd,Schlichting:2025qev}, it is expected to scale
approximately with the cube root of the charged-particle multiplicity,
independently of the absolute system size within this self-similar limit.

Lattice QCD calculations show that at high temperature,
\(p\simeq\epsilon/3\) and the trace anomaly \(\epsilon-3p\) is small,
indicating that QCD is nearly conformal above
\(T\gtrsim3T_c\)~\cite{Borsanyi:2013bia}. This suggests that approximate
conformal scaling may emerge in the quark-gluon plasma created at the LHC,
motivating systematic tests of \(v_2/\varepsilon_2\) as a function of
multiplicity across systems of different sizes. Deviations from the simple
\(N_{\mathrm{ch}}^{1/3}\) scaling, especially in small collision systems,
could signal the breaking of conformality and highlight the role of finite
relaxation times and bulk viscosity in the near-\(T_c\) regime.

\section{\label{sec:analysis} Analysis framework and observables}

Unless stated otherwise, all analyses in this work are carried out in the so-called 
``jet frame''~\cite{Baty:2021ugw}. For each reconstructed jet, a unique coordinate system 
is defined such that the newly defined $z$-axis is aligned with the jet momentum. Both parton-level 
and hadron-level studies are performed in this frame. 

For hadron-level analyses, the momentum vectors of charged hadrons are 
redefined as $\vec{p}^{\,\ast} = (j_T, \eta^\ast, \phi^\ast)$. 
Here, $j_T$ is the transverse momentum relative to the jet axis, while 
$\eta^\ast$ and $\phi^\ast$ denote pseudorapidity and azimuthal angle with 
respect to the jet axis. By construction, $\eta^\ast = 0$ corresponds to a 
transverse direction, while $\eta^\ast \rightarrow \infty$ approaches alignment 
with the jet axis. The approximate boundary of an anti-$k_T$ jet with $R=0.8$ 
lies near $\eta^\ast \simeq 0.86$. The origin of $\phi^\ast$ is defined as 
$\hat{p}_{\text{jet}} \times (\hat{p}_{\text{jet}} \times \hat{z})$, where 
$\hat{p}_{\text{jet}}$ is the unit vector along the jet momentum and 
$\hat{z}$ is the laboratory $+z$ direction. Only relative $\Delta\phi^\ast$ 
values are physically meaningful. 

For parton-level analyses, partons within a cone of radius $R=0.8$ around the 
jet axis are included. Their momentum and position vectors are defined analogously 
to the hadron-level case. 

Jets are generated in $\mathrm{pp}$ collisions at $\sqrt{s}=13\,\mathrm{TeV}$ using \textsc{PYTHIA~8}. After all stages of the transport model, they are reconstructed in the laboratory frame using the anti-$k_T$
algorithm~\cite{Cacciari:2008gp,Cacciari:2011ma} with distance parameter $R=0.8$, 
transverse momentum $\pT^{\text{jet}}>550~\text{GeV}$, and pseudorapidity $|\eta|<1.6$. The jet charged-hadron multiplicity, denoted as \(N_{\mathrm{ch}}^{\mathrm{j}}\), is defined as the number of final charged hadrons clustered into a jet by the algorithm.

\subsection{Initial-state size and geometry}
\label{sec:initial_eccentricity}

The initial-state geometry of the jet system is characterized by the eccentricity vector 
$\bm{\varepsilon_n}$, defined in the transverse coordinate plane ($x^\ast$--$y^\ast$) 
orthogonal to the jet axis ($z^\ast$) in the jet frame:
\begin{align}
\label{eq:ecc_def}
    \bm{\varepsilon_n} &= \varepsilon_n e^{in\Phi_n} \notag \\
    &\equiv \frac{\sum_j w_j e^{in\phi^{\ast,\mathrm{coord}}_j}}{\sum_j w_j},
\end{align}
where $\varepsilon_n$ is the magnitude, $\Phi_n$ the event-plane angle, and 
$\phi^{\ast,\mathrm{coord}}_j$ is the azimuthal angle of the $j^{\rm th}$ constituent 
relative to the energy-weighted centroid:
\begin{equation}
\label{eq:phi_def}
    \phi^{\ast,\mathrm{coord}}_j =
    \operatorname{atan2}\!\left(
        y^\ast_j - \langle y^\ast \rangle_{p},
        x^\ast_j - \langle x^\ast \rangle_{p}
    \right)+\frac{\pi}{2},
\end{equation}
with
\begin{align}
\label{eq:wt_def}
    \langle x^\ast \rangle_{p} &= \frac{\sum_j E_j x^\ast_j}{\sum_j E_j}, &
    \langle y^\ast \rangle_{p} &= \frac{\sum_j E_j y^\ast_j}{\sum_j E_j}, \notag \\
    w_j &= E_j r_j^{\ast 2}, \notag\\
    r_j^{\ast 2} &= (x^\ast_j - \langle x^\ast \rangle_{p})^2 \notag\\
    &\quad +(y^\ast_j - \langle y^\ast \rangle_{p})^2,
\end{align}
where $\langle\cdots\rangle_p$ denotes an energy-weighted average over partons in the jet. For the second-order eccentricity used in this work, the phase shift $\pi/2$ makes the eccentricity vector point along the short axis, i.e., the direction of least energy-weighted spatial extent.

The transverse area of the jet system is estimated from the second moments of constituent positions:
\begin{align}
\label{eq:area_def}
    A_\perp &= \pi 
    \sqrt{
        \langle x^{\ast 2} \rangle_{p} \, \langle y^{\ast 2} \rangle_{p} 
        - \langle x^\ast y^\ast \rangle_{p}^2
    }, \\[2mm]
    \langle x^{\ast 2} \rangle_{p} &= 
    \frac{\sum_j E_j \,(x^\ast_j - \langle x^\ast \rangle_p)^2}{\sum_j E_j}, \\[1mm]
    \langle y^{\ast 2} \rangle_{p} &= 
    \frac{\sum_j E_j \,(y^\ast_j - \langle y^\ast \rangle_p)^2}{\sum_j E_j}, \\[1mm]
    \langle x^\ast y^\ast \rangle_{p} &= 
    \frac{\sum_j E_j \,(x^\ast_j - \langle x^\ast \rangle_p)(y^\ast_j - \langle y^\ast \rangle_p)}{\sum_j E_j}.
\end{align}

These quantities provide complementary information: $\varepsilon_n$ captures shape anisotropy, and $A_\perp$ quantifies overall size.
In addition, \(N_p\), the number of partons inside the jets, indicates how populated the initial transverse area is with partons.

To study the longitudinal structure and test the Bjorken picture in jets, a two-subevent method is employed:
\begin{equation}
\label{eq:epsilon}
    \varepsilon^{a,b}_n \equiv 
    \frac{\langle \bm{\varepsilon_n}(\eta_s^{*a})\,\bm{\varepsilon_{n}}^*(\eta_s^{*b}) \rangle}
         {\sqrt{\langle \bm{\varepsilon_n}(\eta_s^{*b})\,\bm{\varepsilon_n}^*(\eta_s^{*b}) \rangle}}.
\end{equation}
Here, subevents $a$ and $b$ correspond to different (or the same) space-time pseudorapidity 
intervals $\eta_s^{*a}$, $\eta_s^{*b}$ in the jet frame. For $a=b$, Eq.~(\ref{eq:epsilon}) reduces to 
the root-mean-square eccentricity magnitude $\sqrt{\langle \varepsilon_n^2(\eta_s^*)\rangle}$. 
Averages are taken over all jets, weighted by the total weight of each eccentricity vector, 
$W \equiv \sum_j w_j$.

For $a\neq b$, Eq.~(\ref{eq:epsilon}) reflects correlations between separated $\eta_s^*$ intervals. 
To quantify decorrelations explicitly, we define
\begin{equation}
\label{eq:rab_def_vector}
    r^{\varepsilon}_{n}(\eta_s^{*a},\eta_s^{*b}) \equiv 
    \frac{\langle \bm{\varepsilon_n}(\eta_s^{*a})\bm{\varepsilon_n}^*(\eta_s^{*b}) \rangle}
         {\sqrt{\langle \bm{\varepsilon_n}(\eta_s^{*a})\bm{\varepsilon_n}^*(\eta_s^{*a}) \rangle}
          \sqrt{\langle \bm{\varepsilon_n}(\eta_s^{*b})\bm{\varepsilon_n}^*(\eta_s^{*b})\rangle}}.
\end{equation}
$r^{\varepsilon}_{n}=1$ indicates perfect correlation, while deviations below unity signal decorrelation, generally stronger in smaller systems~\cite{Aad:2021ucj,Grosse-Oetringhaus:2024xrl}.

\subsection{Jet substructure observables}

To probe the connection between initial geometry and final-state anisotropies, 
jets are categorized by their substructure using the Soft Drop algorithm~\cite{Larkoski:2014wba} 
implemented in FastJet~\cite{Cacciari:2011ma}. Soft Drop reclusters jet constituents 
with the Cambridge--Aachen algorithm~\cite{Dokshitzer:1997in,Wobisch:1998wt} and removes soft wide-angle radiation until
\begin{equation}
\label{eq:grooming_std}
    \frac{\min(p_{\mathrm{T},i}, p_{\mathrm{T},j})}{p_{\mathrm{T},i}+p_{\mathrm{T},j}} 
    > z_{\mathrm{cut}} \left(\frac{\Delta R_{ij}}{R}\right)^\beta
\end{equation}
is satisfied, where $p_{\mathrm{T},i,j}$ are the subjet transverse momenta, 
$\Delta R_{ij}$ their separation, $R$ the jet radius, and $(z_{\mathrm{cut}},\beta)$ control grooming. 
In this study, $z_{\mathrm{cut}}=0.1$ and $\beta=1$, with variations confirming stability.

Two standard Soft Drop observables are used. The momentum sharing fraction,
\begin{equation}
\label{eq:zg_def}
    z_g = \frac{\min(p_{\mathrm{T},1},p_{\mathrm{T},2})}{p_{\mathrm{T},1}+p_{\mathrm{T},2}},
\end{equation}
quantifies balance of the hard splitting, while the normalized splitting angle,
\begin{equation}
\label{eq:thetag_def}
    \theta_g = \frac{R_g}{R}, \quad 
    R_g = \sqrt{(\eta_1-\eta_2)^2 + (\phi_1-\phi_2)^2},
\end{equation}
characterizes the separation of the two subjets. Jets with large $z_g$ and $\theta_g$ 
are ``two-prong-like,'' reflecting potentially more eccentric initial geometries,
while smaller values correspond to nearly circularly symmetric jets. 
A combined variable $z_g\theta_g^\beta$ is also used to better capture two-prong topologies.

\subsection{Momentum-space anisotropies}

Final-state anisotropies are calculated in the jet frame. The flow vector $\mathbf{Q}_n$~\cite{Luzum:2012da} is
\begin{equation}
\label{eq:q_vector_def}
    \mathbf{Q}_n \equiv Q_n e^{in\Psi_n} =
    \frac{\sum_j w_j e^{in\phi^\ast_j}}{\sum_j w_j},
\end{equation}
where $\phi^\ast_j$ is the azimuthal angle of the $j^\text{th}$ particle in momentum space, 
measured relative to the jet axis, and $\Psi_n$ is the momentum-space event-plane angle. 
The magnitude $Q_n$ quantifies the $n^\text{th}$-order anisotropy. The weights $w_j$
may be the particle transverse momenta or unity; we use unit weights throughout this work.

To probe the response of momentum anisotropies to the initial geometry, we define
\begin{equation}
\label{eq:sab_def_vector}
    v^{s}_n \equiv 
    \frac{\langle \mathbf{Q}_n(\eta^{*a})\,\bm{\varepsilon_{n}}^*(\eta_s^{*b}) \rangle}
         {\sqrt{\langle \bm{\varepsilon_n}(\eta_s^{*b})\,\bm{\varepsilon_n}^*(\eta_s^{*b}) \rangle}}.
\end{equation}
Averages are taken over all jets, weighted by the total weight of each vector, 
$\sum_j w_j$. This correlator quantifies how strongly the final-state momentum 
anisotropy reflects the initial geometric eccentricity.

\subsection{Two-particle $\Delta\eta^*$--$\Delta\phi^*$ correlations}

The two-dimensional (2-D) angular correlation function for particle pairs within jets is calculated as

\begin{equation}
 \frac{1}{N_{\rm ch}^{\rm trg}}\frac{{\rm d}^2N^{\rm pair}}{{\rm d}\Delta\eta^{*} {\rm d}\Delta\phi^{*}}
 = B(0,0)\frac{S(\Delta\eta^{*},\Delta\phi^{*})}{B(\Delta\eta^{*},\Delta\phi^{*})},
\end{equation}

\noindent where $\Delta\eta^*$ and $\Delta\phi^*$ are the relative pseudorapidity and azimuthal angle differences 
in the jet frame for a trigger--associated-particle pair. The signal $S$ and combinatorial background $B$ distributions are defined as

\begin{align}
 S(\Delta\eta^{*},\Delta\phi^{*}) &= \frac{1}{N_{\rm ch}^{\rm trg}}\frac{{\rm d}^2 N^{\rm sig}}{{\rm d}\Delta\eta^{*} {\rm d}\Delta\phi^{*}},\\
 B(\Delta\eta^{*},\Delta\phi^{*}) &= \frac{1}{N_{\rm ch}^{\rm trg}}\frac{{\rm d}^2 N^{\rm combin}}{{\rm d}\Delta\eta^{*} {\rm d}\Delta\phi^{*}}.
\end{align}

\noindent Here, $N_{\rm ch}^{\rm trg}$ is the number of trigger particles in a given $j_T$ range, and $N^{\rm sig}$ and $N^{\rm combin}$ are the numbers of signal and combinatorial pairs, respectively. Trigger and associated particles can be selected from the same or different $j_T$ intervals.

The signal distribution is constructed from particle pairs within the same jet and then averaged over all jets. The combinatorial distribution serves as an uncorrelated reference and accounts for detector acceptance. It is generated by sampling pseudoparticles from the two-dimensional single-particle $(\eta^*,\phi^*)$ distributions for each momentum and multiplicity class across jets. A large number of pseudoparticles $n_{\rm pseudo}$ is generated such that $N^{\rm combin} \approx m N^{\rm sig}$, with $m=10$
, ensuring sufficient statistical precision. Assuming perfect detector efficiency at $(\Delta\eta^*,\Delta\phi^*)=(0,0)$, the bin-by-bin correction is applied via the ratio $B(0,0)/B(\Delta\eta^*,\Delta\phi^*)$.

The 2-D correlation can be further analyzed via a one-dimensional (1-D) Fourier decomposition along the $\Delta\phi^*$ axis:

\begin{equation}
 \frac{1}{N_{\rm ch}^{\rm trg}}\frac{{\rm d}N^{\rm pair}}{{\rm d}\Delta\phi^*} 
 \propto 1 + 2\sum_{n=1}^{\infty} V_{n\Delta}^*(j_T^{\rm trg},j_T^{\rm ass}) \cos(n\Delta\phi^*),
\end{equation}

\noindent where $j_T^{\rm trg}$ and $j_T^{\rm ass}$ are the transverse momenta of trigger and associated particles, respectively. To suppress short-range correlations and isolate long-range effects, the projection is performed over $|\Delta\eta^*|>2$, except in $\Delta\eta^*_\mathrm{min}$-differential $v_2^*$ measurements. The Fourier coefficients are extracted by fitting a three-term cosine series. The single-particle flow coefficients are then obtained via~\cite{Poskanzer:1998yz,Khachatryan:2016txc}:

\begin{equation}
 v_n^*(j_T^{\rm trg}) = \frac{V_{n\Delta}^*(j_T^{\rm trg},j_T^{\rm ass})}{\sqrt{V_{n\Delta}^*(j_T^{\rm ass},j_T^{\rm ass})}},
\end{equation}
where both $V_{n\Delta}^*$ and $v_n^*$ are extracted in the jet frame.

\section{\label{sec:results} Results}
\subsection{Initial-state geometry of jets}

\begin{figure*}[t]
    \centering
    \includegraphics[width=\linewidth]{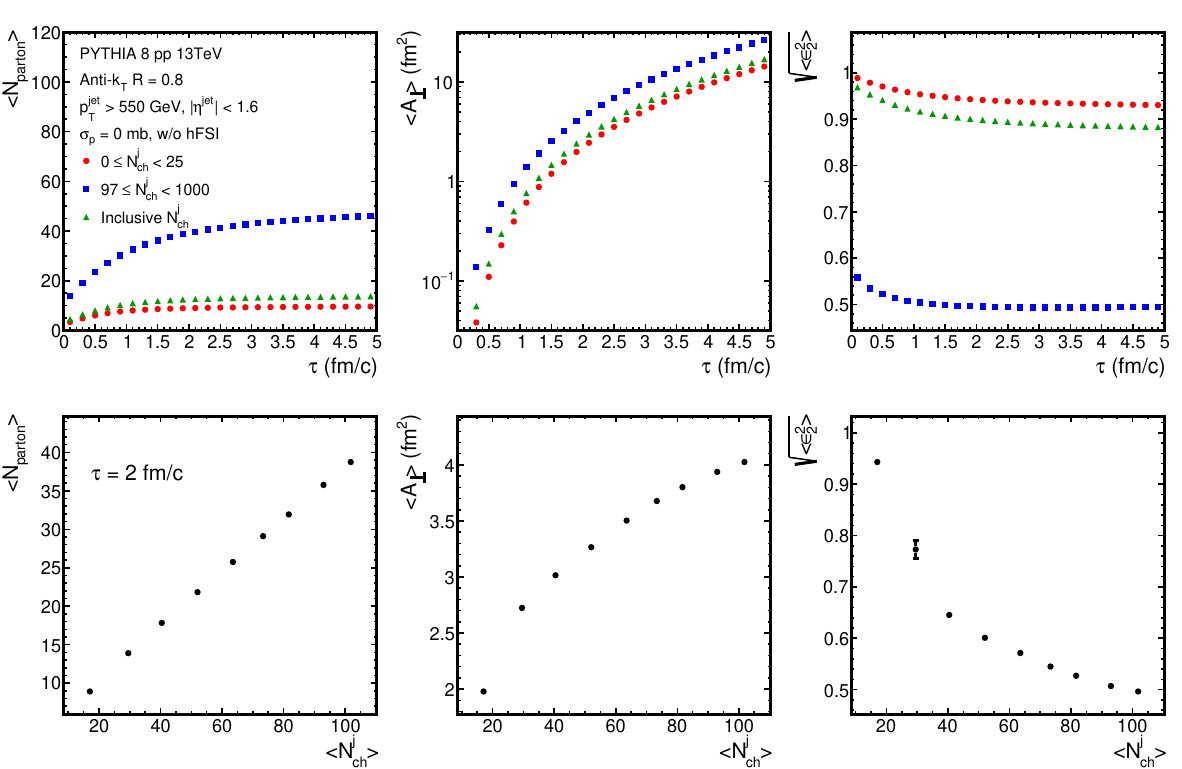}
    \caption{Time evolution and multiplicity dependence of the initial-state geometry of jets in the jet frame for $\sigma_{p}=0$ mb without hadronic final-state interactions. Upper panels: average parton number $\langle N_{\mathrm{parton}} \rangle$, transverse area $\langle A_{\perp} \rangle$, and RMS eccentricity $\sqrt{\langle \varepsilon_{2}^{2} \rangle}$ as functions of proper time $\tau$ for low-multiplicity, high-multiplicity, and inclusive jets. Lower panels: the same observables as functions of $\langle N_{\mathrm{ch}}^{j} \rangle$ at $\tau = 2$ fm/$c$.}
    \label{fig:basic}
\end{figure*}

Figure~\ref{fig:basic} establishes the basic features of the initial transverse geometry, characterized by the parton number, transverse area, and eccentricity.
The upper panels show the time evolution of the average parton number, transverse area, and root-mean-square (RMS) eccentricity for jets in different multiplicity classes; \(\tau\) denotes the proper time of each parton. In all \(N_{\mathrm{ch}}^j\) classes, the average parton number increases with \(\tau\) and reaches a plateau at approximately $\tau=2\,\mathrm{fm}/c$. The transverse area also grows rapidly, indicating rapid transverse expansion during the early stage of jet formation. The eccentricity decreases during this stage and stabilizes at approximately $1\,\mathrm{fm}/c$, possibly because the partons free-stream after production.
The lower panels show these observables as functions of jet multiplicity at $\tau=2\,\mathrm{fm}/c$, when parton formation is largely complete. The parton number is proportional to the jet multiplicity, as expected for high-multiplicity jets.

\begin{figure*}[tb]
    \centering
    \includegraphics[width=\linewidth]{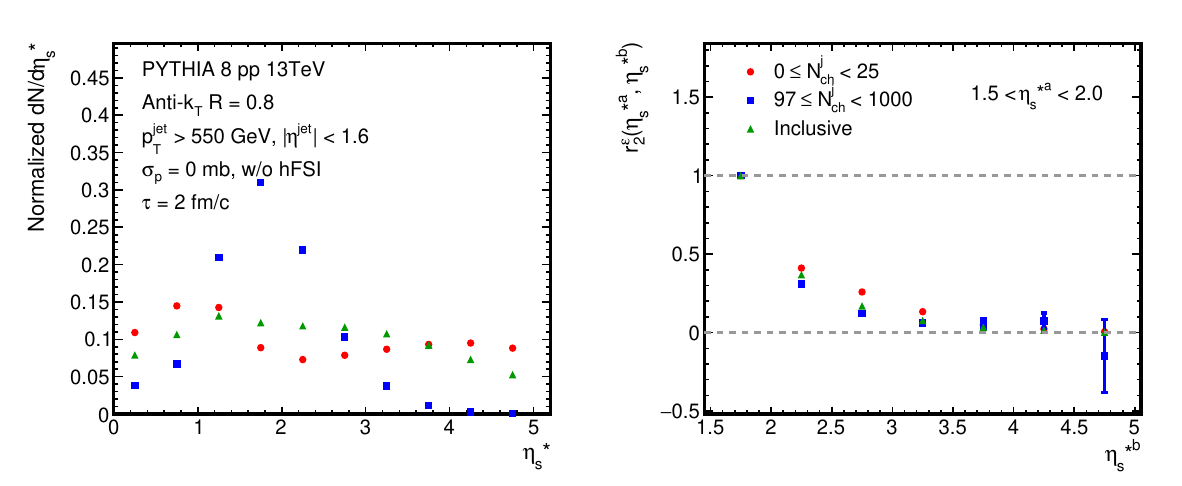}
    \caption{Longitudinal structure of the jet initial state in the jet frame for $\sigma_{p}=0$ mb without hadronic final-state interactions. Left: normalized distribution of the parton space-time pseudorapidity $\eta_{s}^{*}$ at $\tau = 2$ fm/$c$. Right: eccentricity correlation $r_{2}^{\epsilon}(\eta_{s}^{*a},\eta_{s}^{*b})$ as a function of $\eta_{s}^{*b}$, with the reference interval $1.5<\eta_{s}^{*a}<2.0$, for different jet-multiplicity classes.}
    \label{fig:etas}
\end{figure*}

To illustrate the initial longitudinal structure of jets, Fig.~\ref{fig:etas} presents the parton space-time pseudorapidity distributions and the correlations between eccentricities in different space-time pseudorapidity intervals, as defined in Eq.~\ref{eq:rab_def_vector}.
The distribution for high-multiplicity jets shifts toward smaller \(\eta_s^*\) relative to that for low-multiplicity jets, suggesting larger parton-emission angles with respect to the jet axis.
For the correlation in the right panel, the interval \(1.5<\eta_s^*<2.0\) is used as the reference and correlated with the other intervals. The correlation decreases rapidly as the rapidity gap increases, indicating stronger decorrelation than in larger systems. However, it remains nonzero up to a rapidity gap of at least 2, suggesting that a Bjorken-like picture may remain qualitatively useful even in such a small system.

\subsection{Jet-substructure engineering of the initial geometry}

\begin{figure*}[t]
    \centering
    \includegraphics[width=\linewidth]{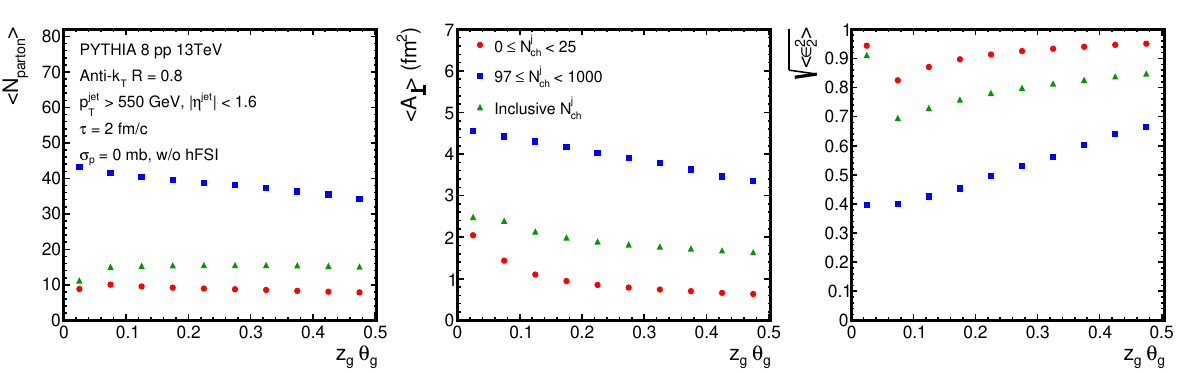}
    \caption{Dependence of the initial-state jet geometry on the substructure variable $z_{g}\theta_{g}$ at $\tau = 2$ fm/$c$ for $\sigma_{p}=0$ mb without hadronic final-state interactions. Shown are the average parton number, transverse area, and RMS eccentricity for low-multiplicity, high-multiplicity, and inclusive jets.}
    \label{fig:zgtg}
\end{figure*}

The correlation between final-state anisotropy and initial-state geometry is central to understanding the origin of collective flow. Because the initial geometry cannot be measured directly, an important question is whether experimentally accessible jet substructure can serve as its proxy. Jets with large \(z_g\) and \(\theta_g\) have a more pronounced two-prong structure and may therefore have larger eccentricities, whereas jets with smaller values may be more nearly azimuthally symmetric. If this relation holds, the combined variable \(z_g\theta_g\) can be used to select jets with different eccentricities and thereby engineer the initial geometry.
Figure~\ref{fig:zgtg} shows the dependence of the initial-state observables on \(z_g\theta_g\). The parton number and transverse area depend only weakly on \(z_g\theta_g\), whereas the eccentricity is generally positively correlated with it. This trend is particularly evident for the high-multiplicity jets of primary interest and demonstrates the feasibility of engineering the initial geometry through jet substructure.

\subsection{Final-state anisotropies and their origins}

\begin{figure}[tb]
    \centering
    \includegraphics[width=\linewidth]{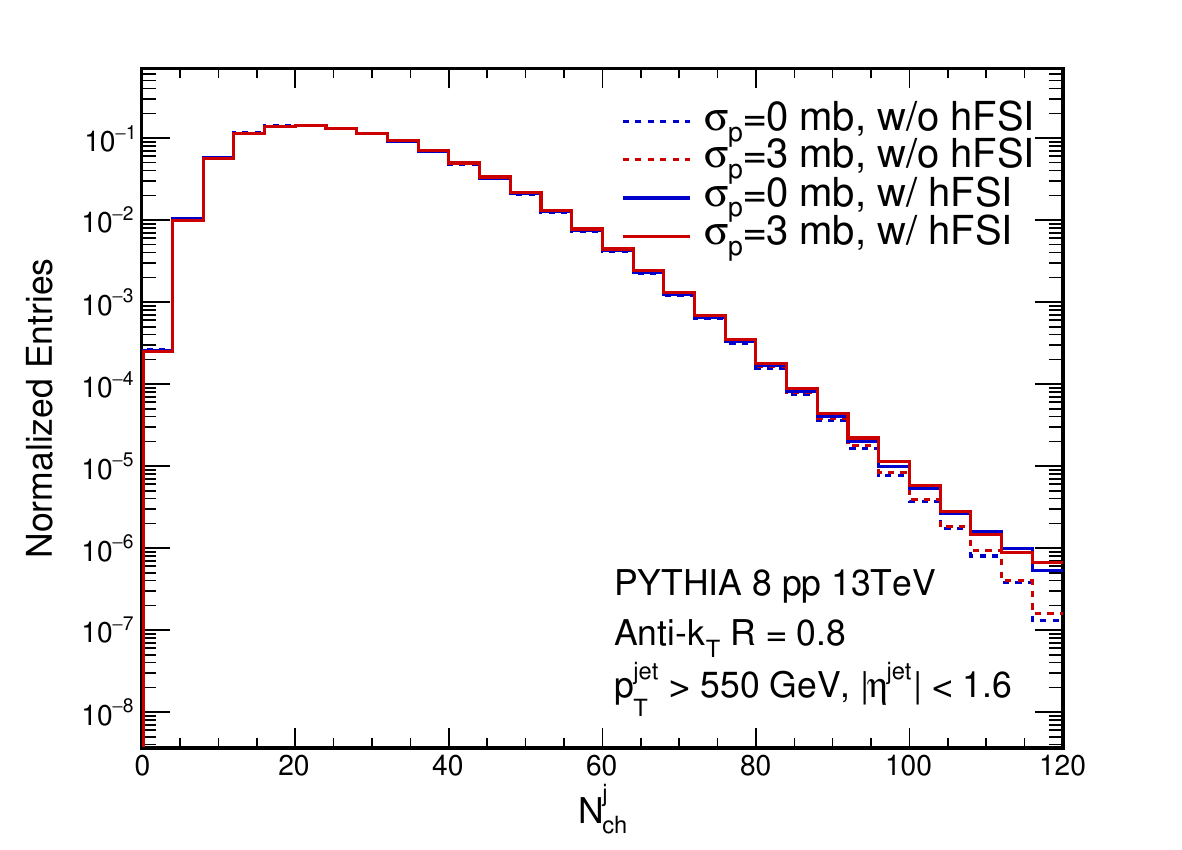}   
    \caption{Jet charged-hadron multiplicity distributions for different final-state interaction settings. Results are shown for $\sigma_{p}=0$ and $3$ mb, each with and without hadronic final-state interactions.}
    \label{fig:nch}
\end{figure}

We study final-state anisotropies and their correlations with the initial geometry. The partonic rescattering strength is controlled by the cross section \(\sigma_p\) in ZPC, while hadronic final-state interactions (hFSI) can be enabled or disabled in UrQMD. These variations separate the contributions of the two interaction stages.

\begin{figure}[tb]
    \centering
    \includegraphics[width=\linewidth]{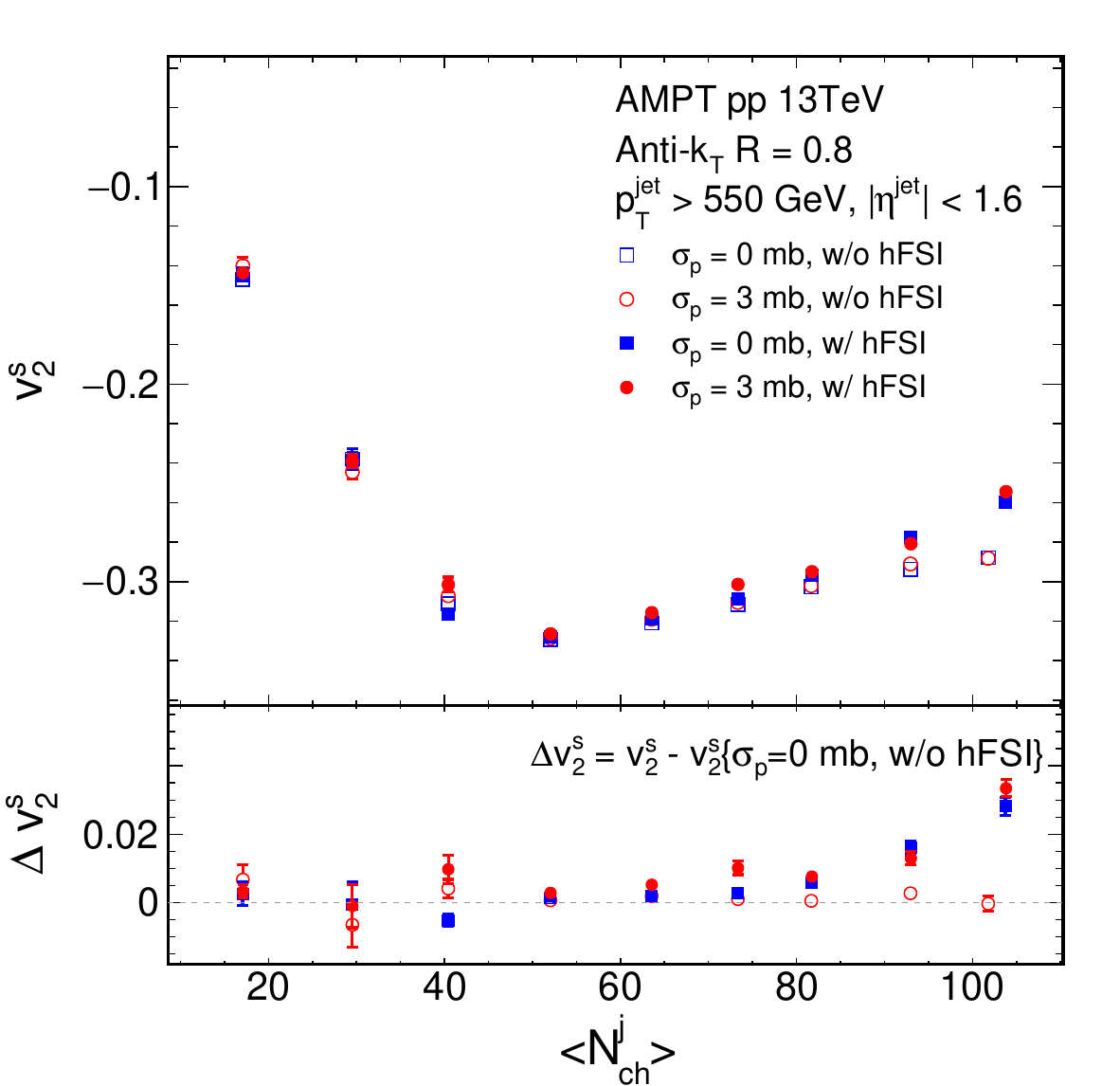}
    \caption{Geometry-response correlator $v_{2}^{s}$ as a function of jet multiplicity for different final-state interaction settings. Upper panel: $v_{2}^{s}$ versus $\langle N_{\mathrm{ch}}^{j} \rangle$. Lower panel: $\Delta v_{2}^{s}$ relative to the $\sigma_{p}=0$ mb baseline without hFSI.}
    \label{fig:v2sVsNch}
\end{figure}

Figure~\ref{fig:nch} compares the \(N_{\mathrm{ch}}^{\mathrm{j}}\) distributions for the different final-state interaction settings. Although all cases have a similar overall shape, hFSI shifts the distributions toward higher multiplicities, whereas \(\sigma_p\) has a negligible effect. Inelastic hadronic interactions can produce additional hadrons, while the partonic scattering implemented in this model is purely elastic and therefore has little effect on the hadron multiplicity. The shift in the characteristic high-multiplicity region is modest---only a few charged hadrons---indicating that hFSI has a relatively small effect on the \(N_{\mathrm{ch}}^{\mathrm{j}}\) distributions.

\begin{figure}[tb]
    \centering
    \includegraphics[width=\linewidth]{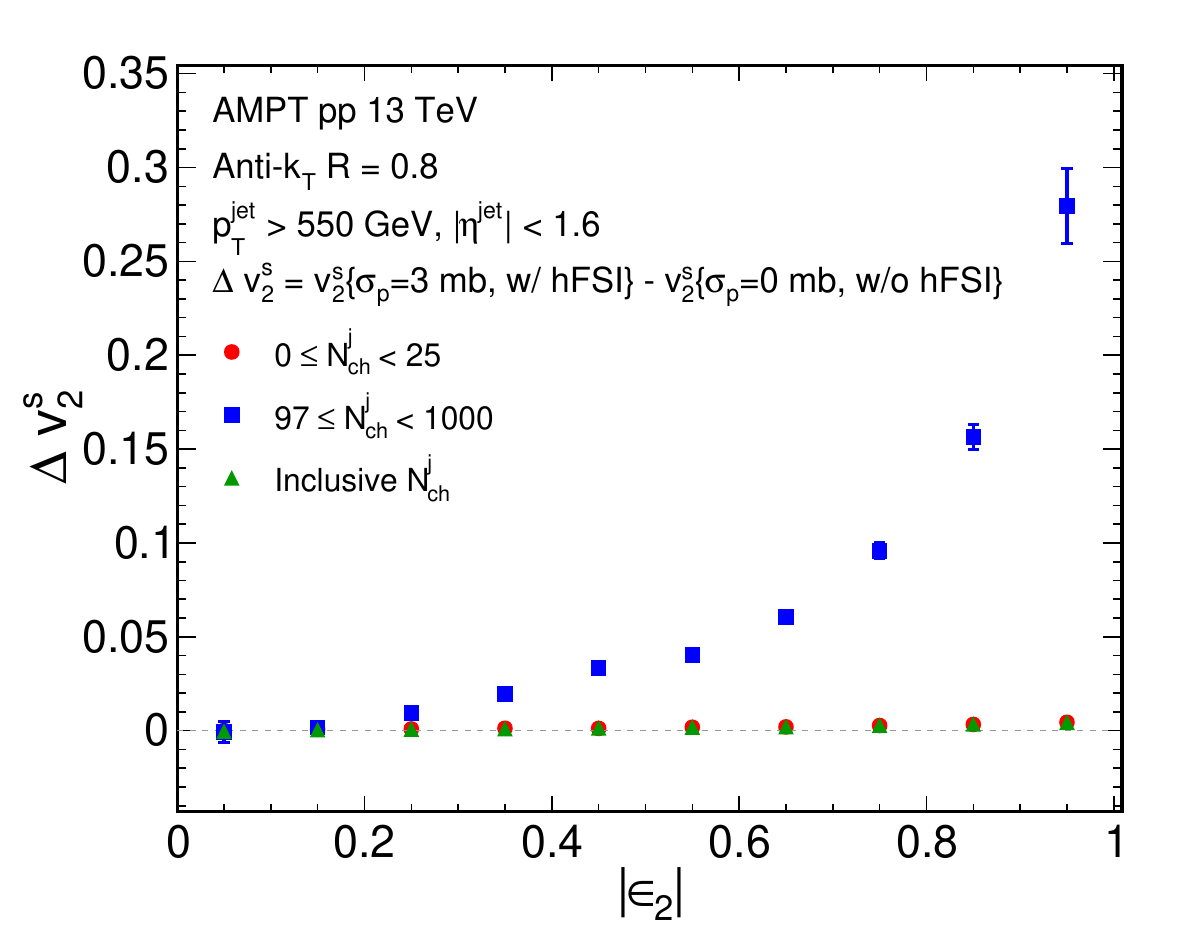}
    \caption{$\Delta v_{2}^{s}$ as a function of the initial eccentricity magnitude $|\varepsilon_{2}|$. The excess is defined relative to the $\sigma_{p}=0$ mb baseline without hFSI and is shown for $\sigma_{p}=3$ mb with hFSI in different multiplicity classes.}
    \label{fig:Dv2sVsEcc}
\end{figure}

The observable \(v_2^s\), defined in Eq.~\ref{eq:sab_def_vector} for $n=2$, quantifies the response of the final-state momentum anisotropy to the initial-state geometric anisotropy. Because Fig.~\ref{fig:etas} shows strong decorrelation between different pseudorapidity intervals, we apply no pseudorapidity gap in the \(v_2^s\) analysis. We use the case without final-state interactions ($\sigma_p=0\,\mathrm{mb}$ without hFSI) as a baseline to subtract the nonflow contribution.
The upper panel of Fig.~\ref{fig:v2sVsNch} shows the multiplicity dependence of $v_2^s$. The values are predominantly negative, which follows directly from the definition of the azimuthal angle in Eq.~(\ref{eq:phi_def}), where an additional phase of $+\pi/2$ is introduced such that the eccentricity vector points along the ``short-axis'' of the initial geometry. In this convention, the geometry-induced anisotropy gives a negative $v_2^s$.
The lower panel presents the enhancement, $\Delta v_2^s$, induced by final-state interactions. At low multiplicity, $\Delta v_2^s$ is consistent with zero, indicating little modification from rescatterings. As the jet multiplicity increases, a positive $\Delta v_2^s$ develops. This enhancement arises almost entirely from hadronic rescatterings, while partonic rescatterings have a negligible effect within the present model setup.

To illustrate the dependence of the geometry response on the initial geometric anisotropy, Fig.~\ref{fig:Dv2sVsEcc} presents \(\Delta v_2^s\) as a function of eccentricity for $\sigma_p=3\,\mathrm{mb}$ with hFSI. The response grows rapidly with eccentricity, indicating stronger flow responses in more eccentric jets.

\begin{figure}[tb]
    \centering
    \includegraphics[width=\linewidth]{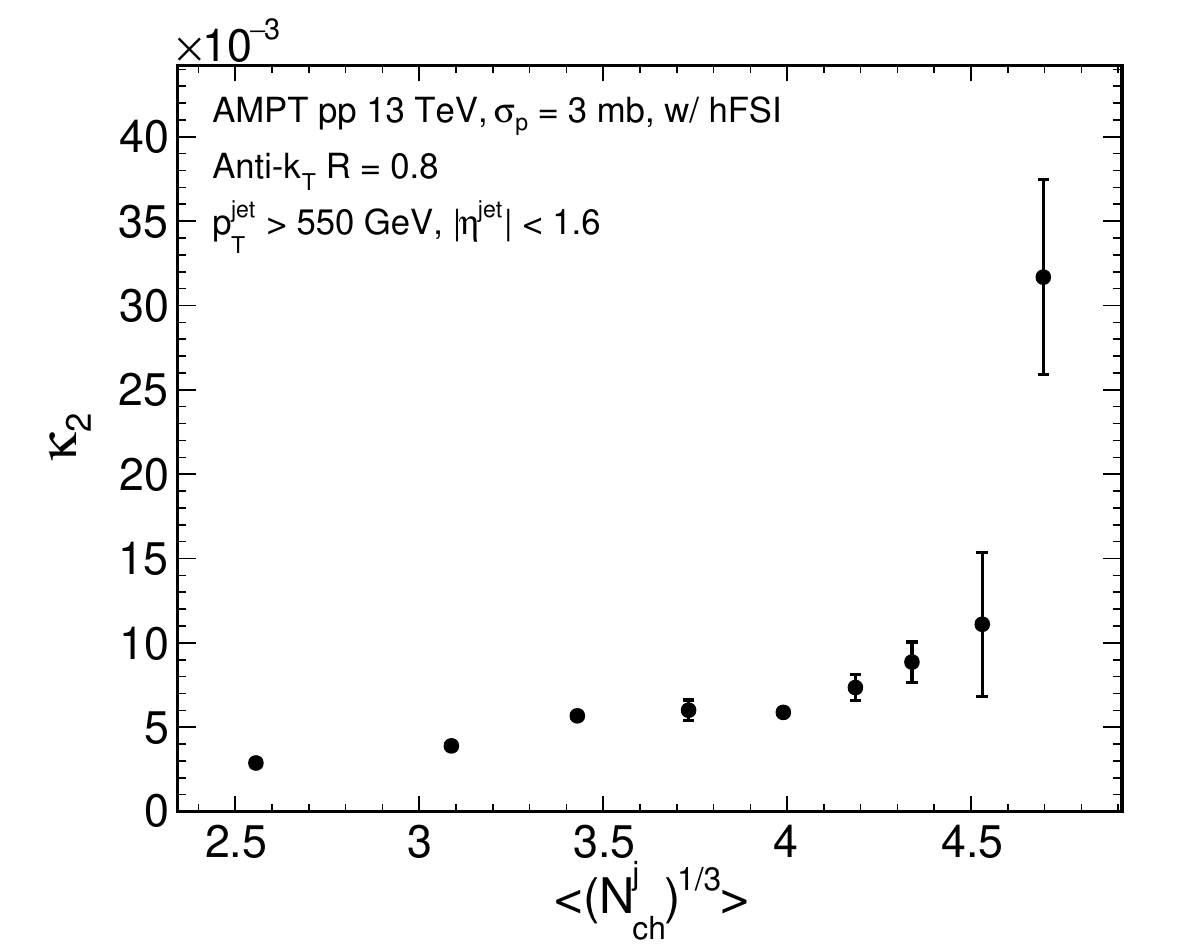}
    \caption{Response coefficient $\kappa_{2}$ extracted from fits to $\Delta v_{2}^{s}(\epsilon_{2})$ as a function of the cube root of jet multiplicity for the $\sigma_{p}=3$ mb with hFSI case.}
    \label{fig:k2VsNch}
\end{figure}

The response coefficients can be extracted from the distributions in Fig.~\ref{fig:Dv2sVsEcc}. Because nonlinear effects are sizable, we fit the distributions with~\cite{Noronha-Hostler:2015dbi}
\begin{equation}
    \Delta v_2^s = \kappa_2 \varepsilon_2 + \kappa_2^{\prime} |\varepsilon_2|^2 \varepsilon_2,
\label{eq:fit_function}
\end{equation}
where \(\kappa_2\) and \(\kappa_2^\prime\) are the linear and cubic response coefficients, respectively. Guided by the conformal scaling relation derived in Sec.~\ref{sec:opacity_scaling}, Fig.~\ref{fig:k2VsNch} presents the fitted \(\kappa_2\) as a function of \(\langle(N_{\mathrm{ch}}^{\mathrm{j}})^{1/3}\rangle\). The coefficient increases without an indication of the saturation reported in Ref.~\cite{Schlichting:2025qev}; qualitatively, this behavior is compatible with the dilute-response regime suggested by the small partonic opacity discussed below.

\begin{figure}[tb]
    \centering
    \includegraphics[width=\linewidth]{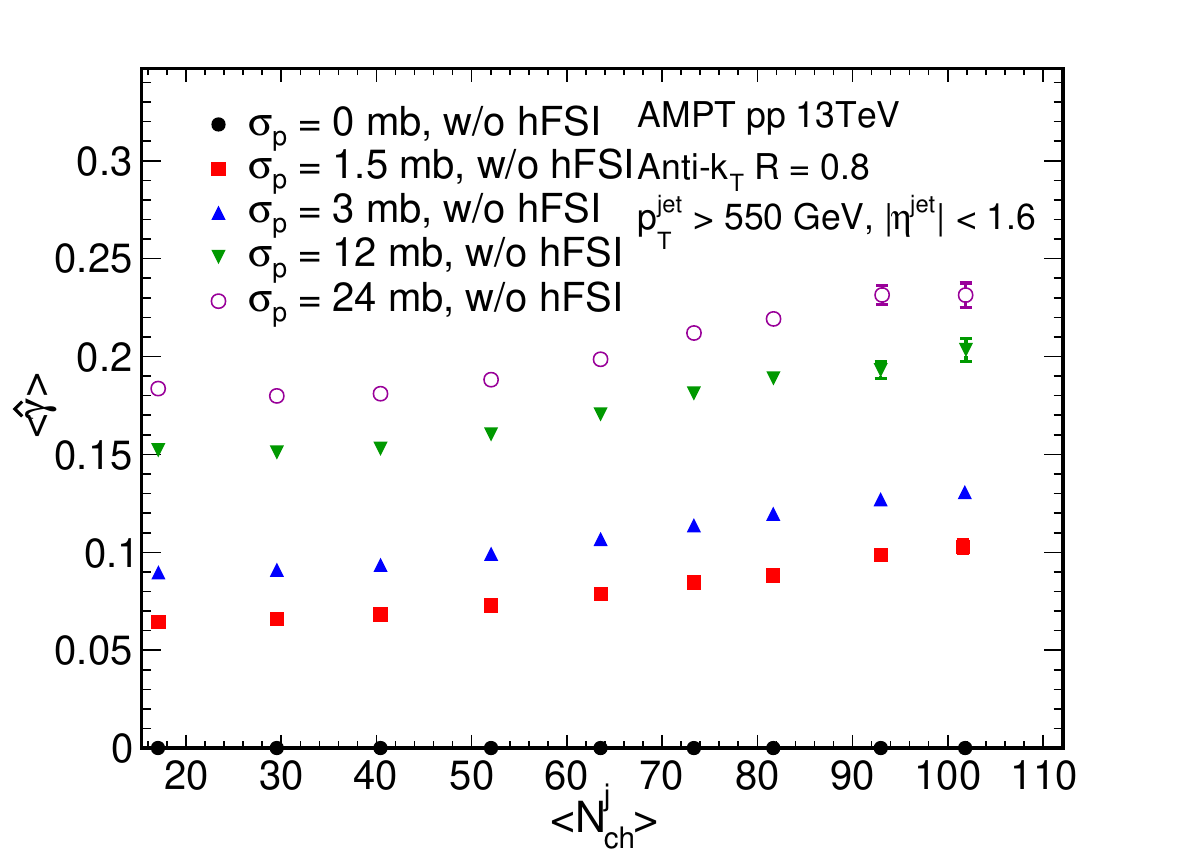}
    \caption{Average partonic opacity as a function of jet multiplicity for different partonic cross sections in ZPC.}
    \label{fig:opacity}
\end{figure}

The opacity defined in Eq.~\ref{eq:opacity} can be estimated directly from the average number of scatterings experienced by each parton and quantifies the interaction strength in the medium. Because the corresponding information is unavailable from UrQMD, we examine the partonic opacity in ZPC to obtain a qualitative picture. Figure~\ref{fig:opacity} shows the partonic opacity as a function of jet multiplicity for several values of \(\sigma_p\). The opacity is larger in high-multiplicity jets and for larger \(\sigma_p\). However, even for the phenomenologically large default value \(\sigma_p=3\,\mathrm{mb}\), the opacity is only approximately 0.1. This small opacity may explain the negligible role of partonic scattering in the final-state anisotropy.

\begin{figure}[tb]
    \centering
    \includegraphics[width=\linewidth]{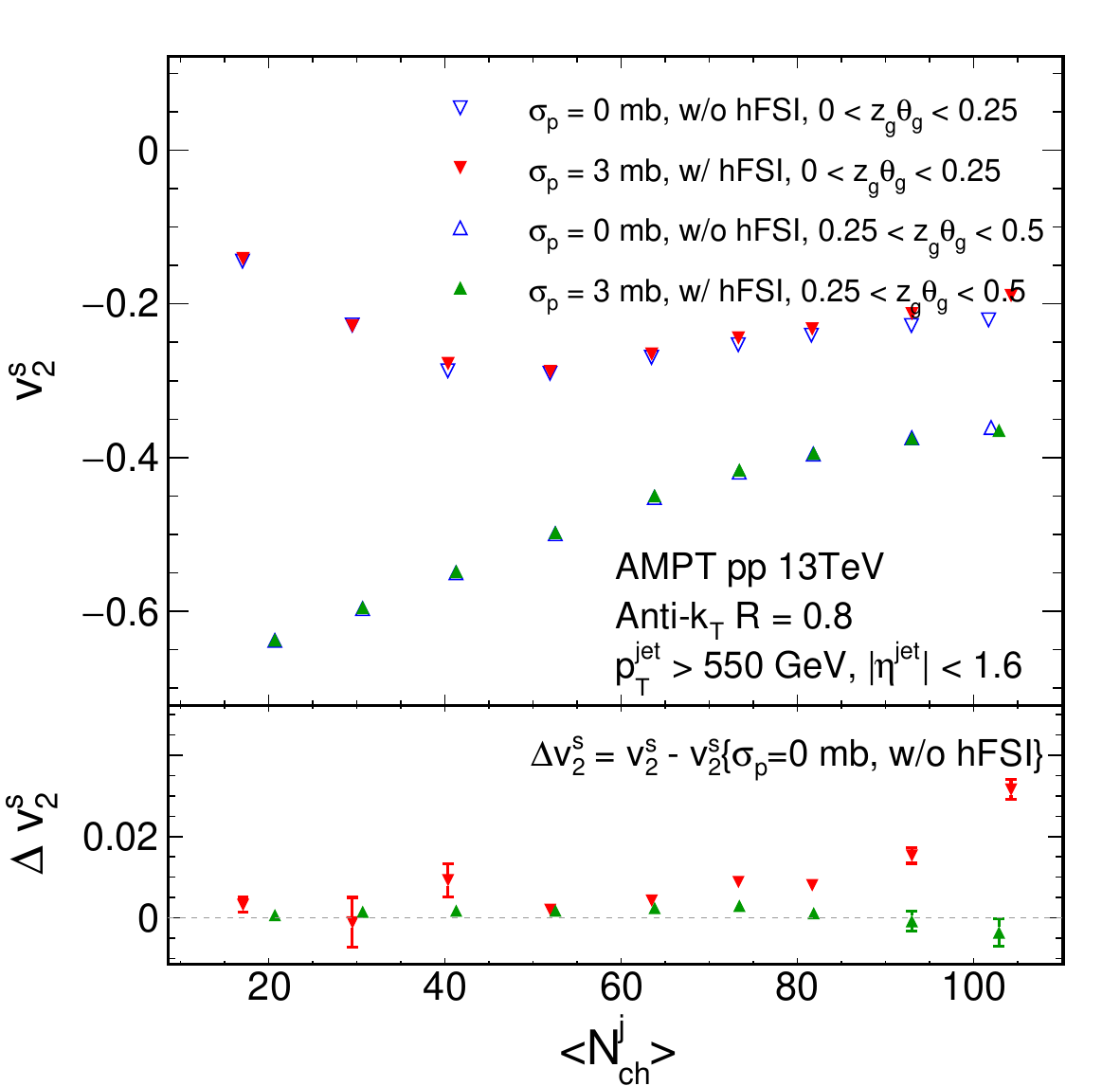}
    \caption{Substructure dependence of the geometry-response correlator $v_{2}^{s}$. Upper panel: $v_{2}^{s}$ versus $\langle N_{\mathrm{ch}}^{j} \rangle$ for different $z_{g}\theta_{g}$ classes, with and without final-state interactions. Lower panel: corresponding excess $\Delta v_{2}^{s}$ relative to the no-FSI baseline. The enhancement is larger for smaller $z_{g}\theta_{g}$ jets.}
    \label{fig:v2sVsNch_zgtg}
\end{figure}

\begin{figure}[tb]
    \centering
    \includegraphics[width=\linewidth]{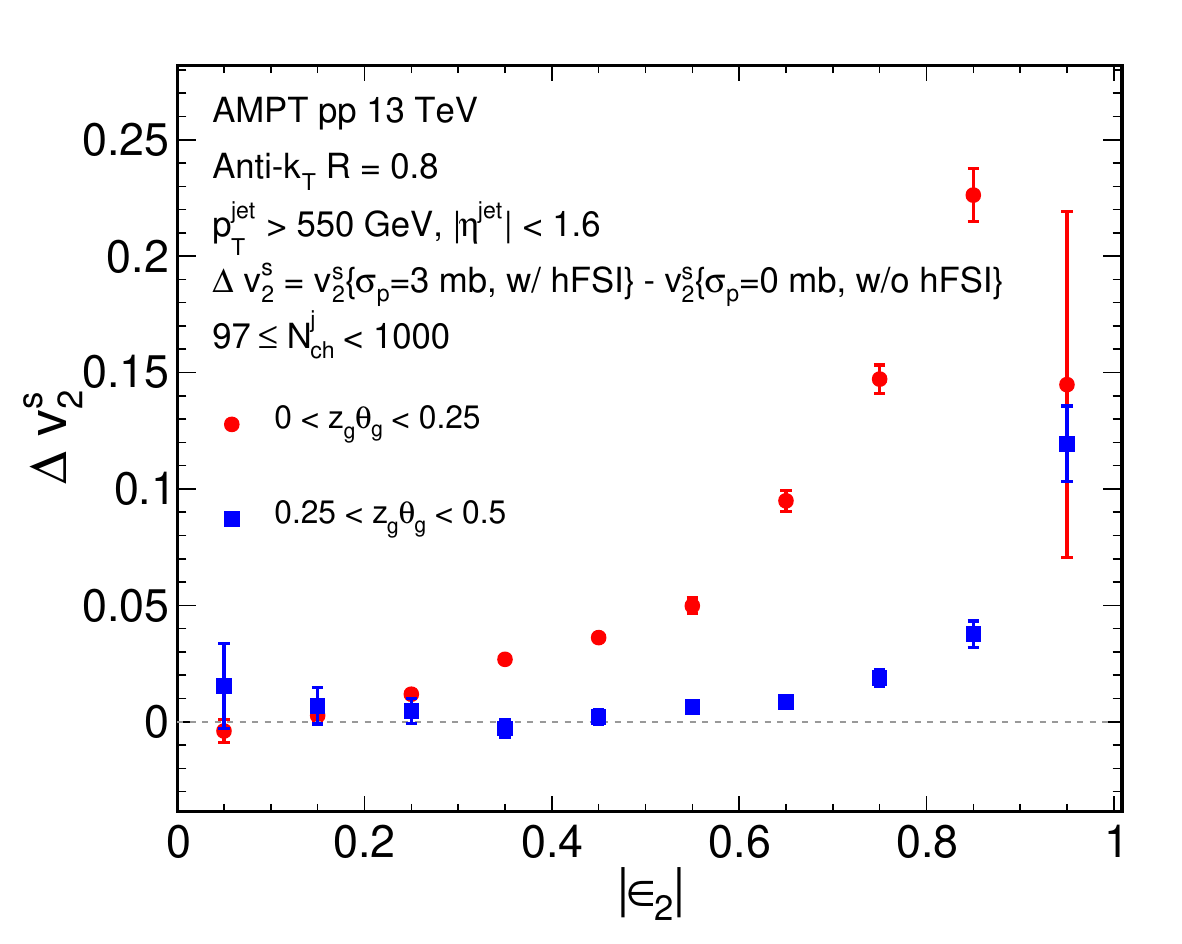}
    \caption{Excess geometry-response correlator $\Delta v_{2}^{s}$ versus initial eccentricity $|\epsilon_{2}|$ for different $z_{g}\theta_{g}$ classes in high-multiplicity jets.}
    \label{fig:Dv2sVsEcc_zgtg}
\end{figure}

\begin{figure}[tb]
    \centering
    \includegraphics[width=\linewidth]{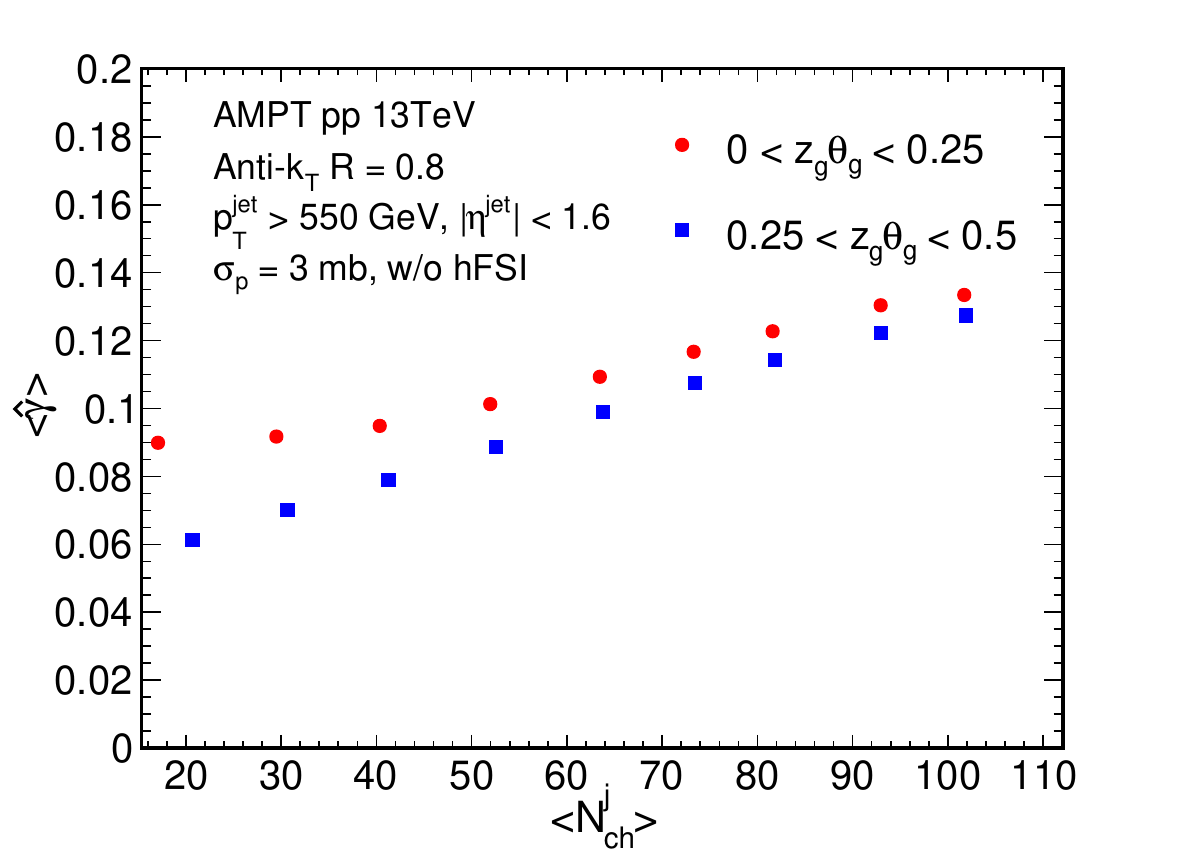}
    \caption{Substructure dependence of the average partonic opacity for $\sigma_{p}=3$ mb without hadronic final-state interactions.}
    \label{fig:opacity_zgtg}
\end{figure}

Figures~\ref{fig:v2sVsNch_zgtg}--\ref{fig:opacity_zgtg} present the corresponding results after the jets are separated according to their Soft Drop substructure. Figure~\ref{fig:v2sVsNch_zgtg} shows that the enhancement \(\Delta v_2^s\) is consistently larger for one-prong jets than for two-prong jets. Nevertheless, Fig.~\ref{fig:Dv2sVsEcc_zgtg} demonstrates that \(\Delta v_2^s\) increases monotonically with the initial eccentricity in both categories, indicating that the geometry--response correlation persists after the substructure selection. Figure~\ref{fig:opacity_zgtg} further shows that one-prong jets have systematically larger transport opacity. This larger opacity is consistent with their stronger \(\Delta v_2^s\), suggesting that the final-state anisotropy is governed by both the initial geometry and the strength of subsequent rescattering.

\subsection{Two-particle correlations}

\begin{figure*}[t]
    \centering
    \includegraphics[width=0.48\textwidth]{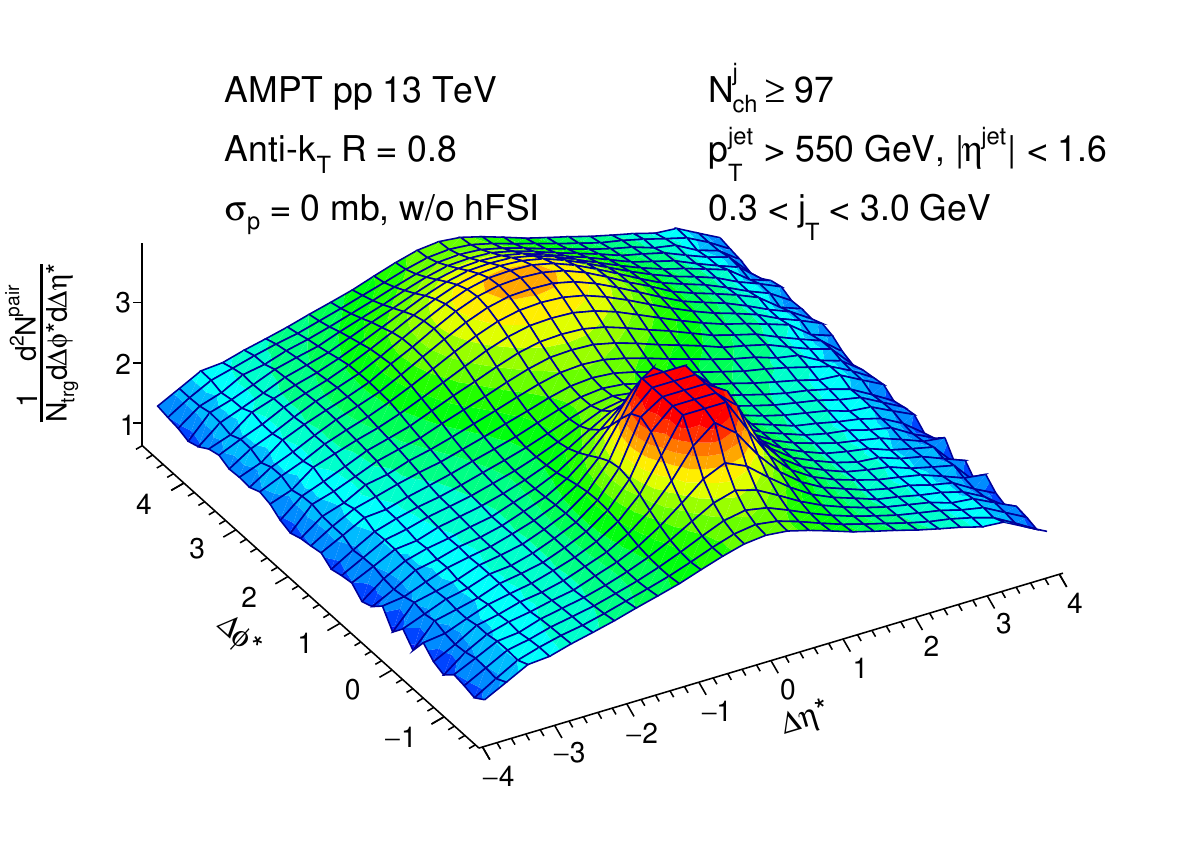}
    \hfill
    \includegraphics[width=0.48\textwidth]{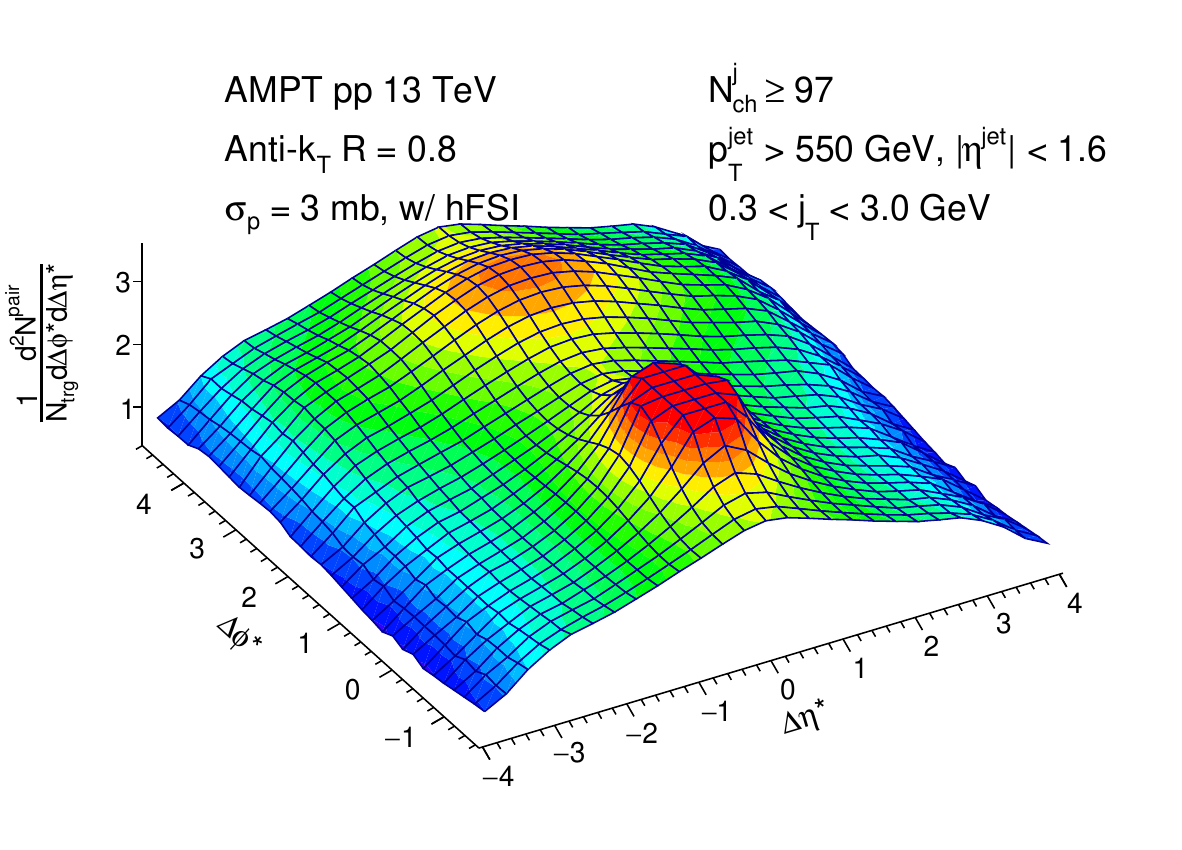}
    \caption{Two-dimensional $\Delta\eta^{*}$--$\Delta\phi^{*}$ correlation functions in high-multiplicity jets for charged hadrons with $0.3<j_{T}<3.0$ GeV. Left: $\sigma_{p}=0$ mb without hFSI. Right: $\sigma_{p}=3$ mb with hFSI.}
    \label{fig:2D_correlation}
\end{figure*}

\begin{figure*}[t]
    \centering
    \includegraphics[width=0.48\textwidth]{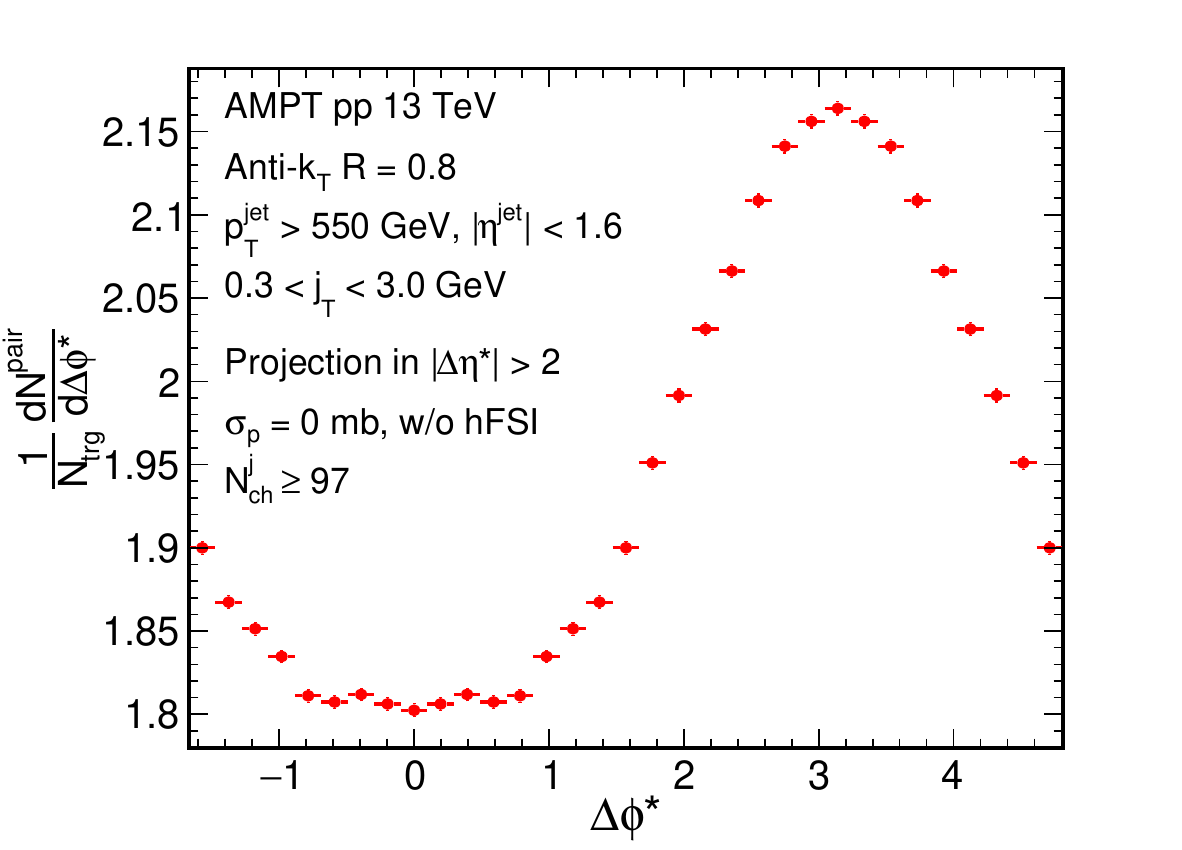}
    \hfill
    \includegraphics[width=0.48\textwidth]{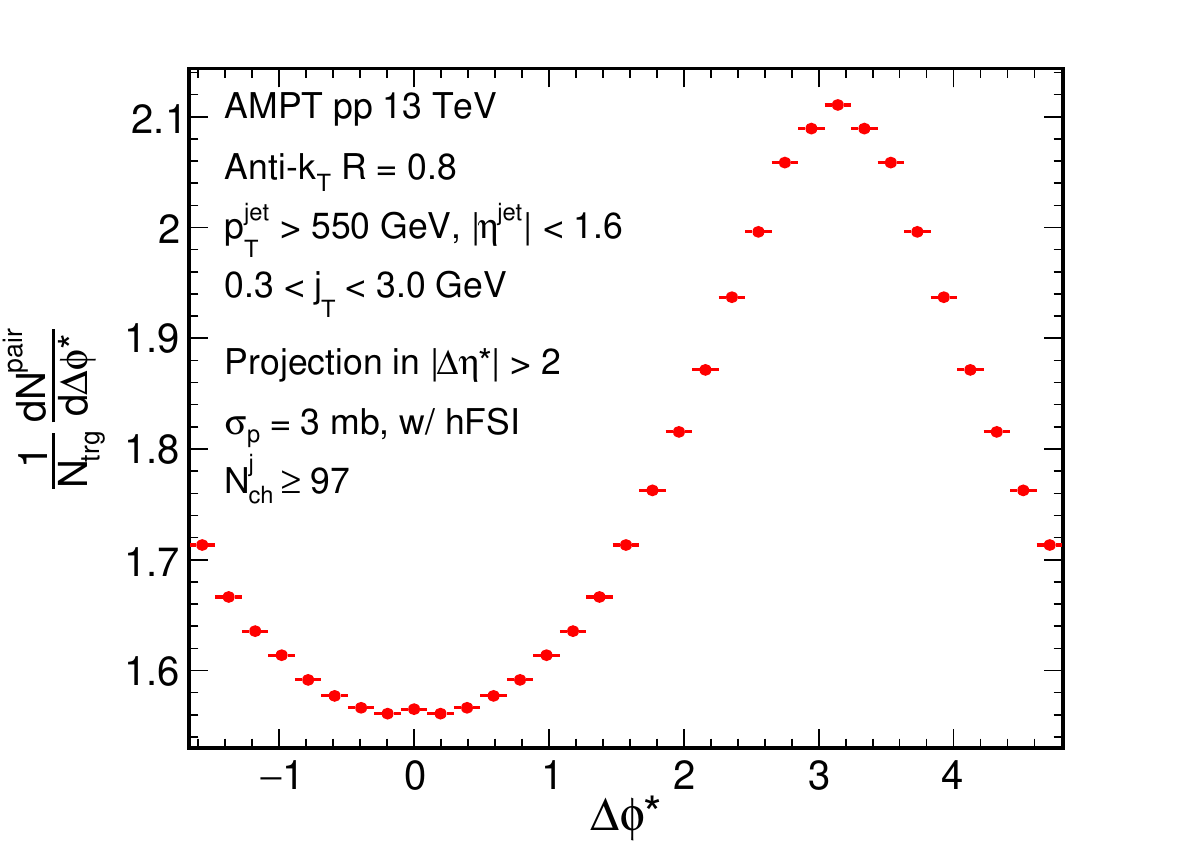}
    \caption{One-dimensional $\Delta\phi^{*}$ correlation functions with $|\Delta\eta^{*}|>2$. Left: $\sigma_{p}=0$ mb without hFSI. Right: $\sigma_{p}=3$ mb with hFSI.}
    \label{fig:1D_correlation}
\end{figure*}

We now assess how well the model reproduces the experimental observations and discuss the implications for future measurements. Figure~\ref{fig:2D_correlation} presents the two-dimensional $\Delta\eta^*$--$\Delta\phi^*$ correlation functions for high-multiplicity jets with and without final-state interactions. The overall structure resembles that observed in heavy-ion collisions. Although no obvious long-range near-side ridge is resolved directly in the two-dimensional distribution, the dominant short-range near-side peak can obscure a weaker long-range correlation. We therefore project the distributions onto $\Delta\phi^*$ for $|\Delta\eta^*|>2$, as shown in Fig.~\ref{fig:1D_correlation}. A slight enhancement near $\Delta\phi^*=0$ becomes visible when final-state interactions are enabled.

 \begin{figure}[tb]
    \centering
    \includegraphics[width=\linewidth]{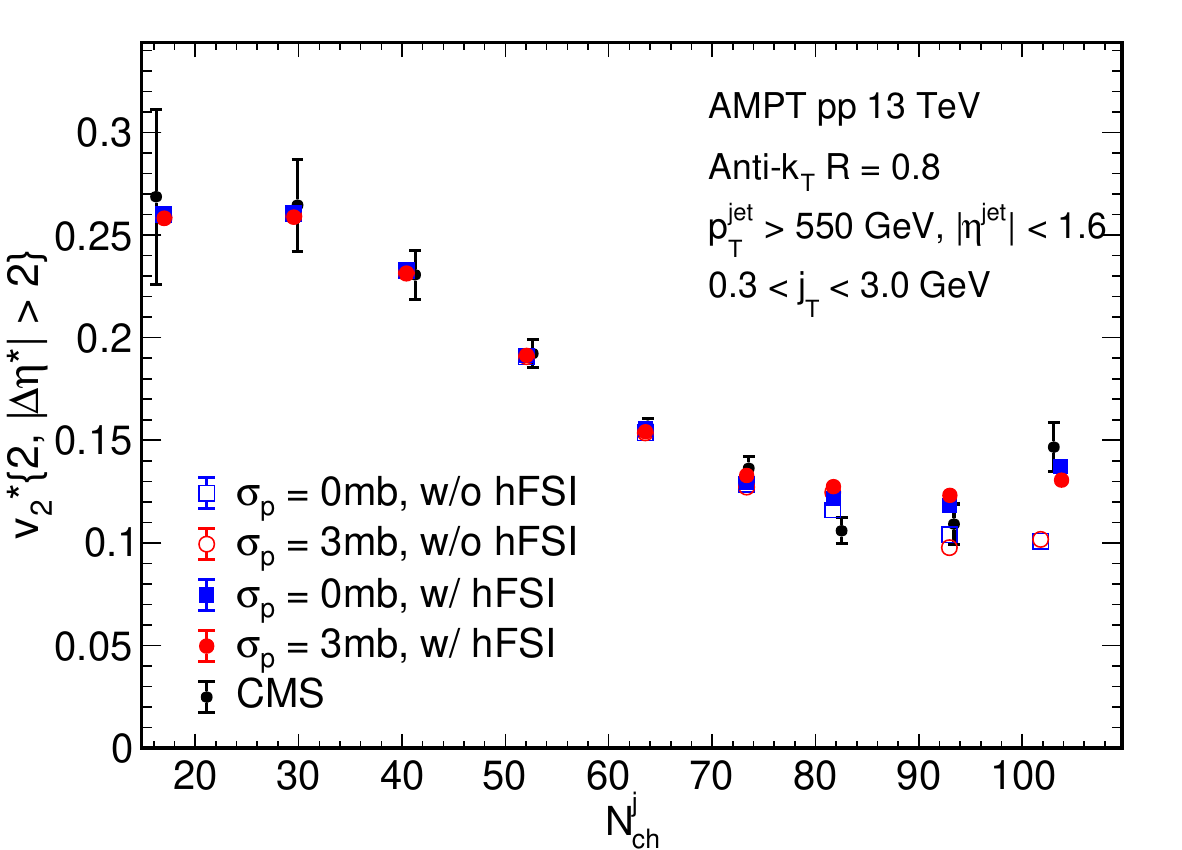}
    \caption{$v_{2}^{*}\{2,|\Delta\eta^{*}|>2\}$ as a function of jet multiplicity, compared with the CMS measurement.}
    \label{fig:v2}
\end{figure}

The value of $v_2^*$ extracted from the one-dimensional $\Delta\phi^*$ correlation functions is presented in Fig.~\ref{fig:v2}. Hadronic final-state interactions produce a clear enhancement, bringing the model into agreement with the CMS measurement within uncertainties. In contrast, partonic scattering has a negligible effect, consistent with the geometry--response analysis above.

 \begin{figure}[tb]
    \centering
    \includegraphics[width=\linewidth]{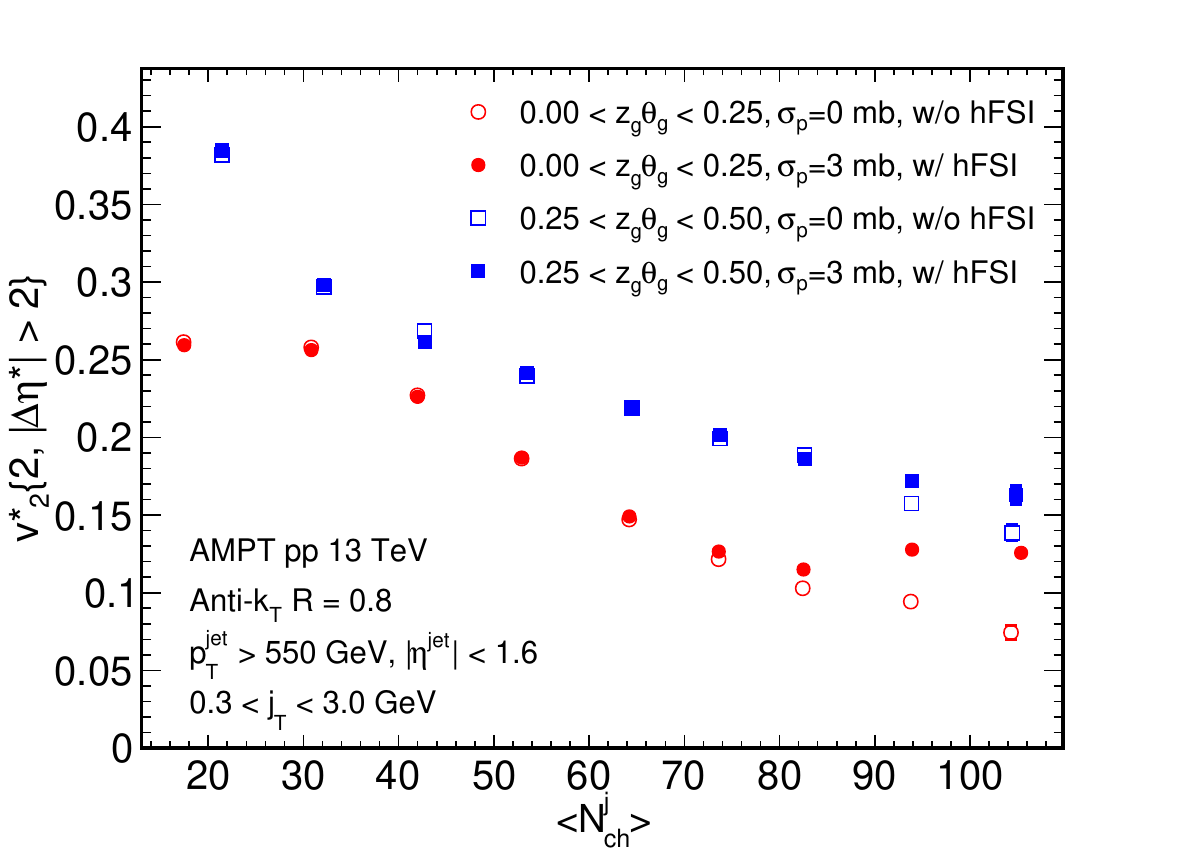}
    \caption{Substructure dependence of $v_{2}^{*}\{2,|\Delta\eta^{*}|>2\}$ as a function of jet multiplicity.}
    \label{fig:v2_zgtg}
\end{figure}

Figure~\ref{fig:v2_zgtg} shows \(v_2^*\) for different \(z_g\theta_g\) classes. Both classes exhibit an enhancement relative to the baseline when final-state interactions are included. The enhancement is more pronounced for jets with smaller \(z_g\theta_g\), despite their smaller average initial eccentricity, consistent with the \(v_2^s\)-based analysis.

 \begin{figure}[tb]
    \centering
    \includegraphics[width=\linewidth]{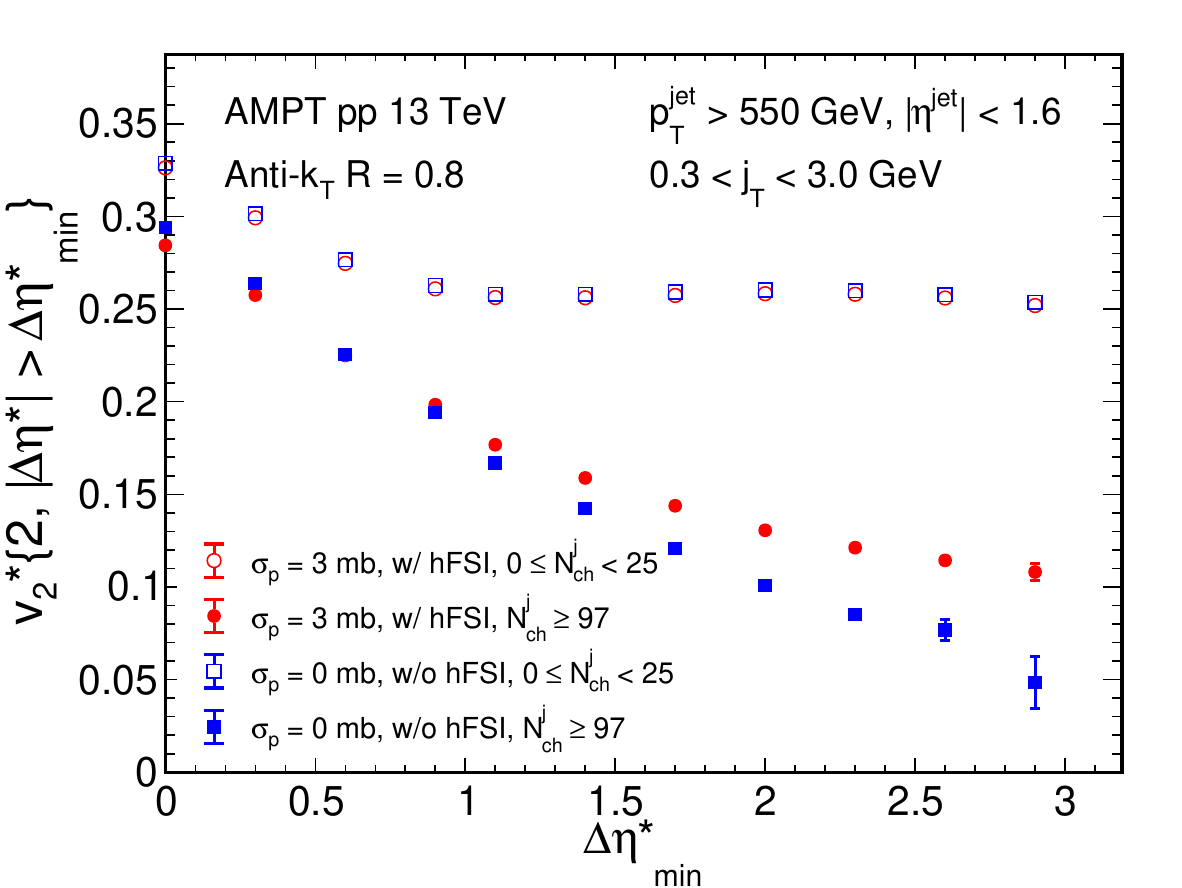}
    \caption{Pseudorapidity-gap dependence of $v_{2}^{*}$ for low- and high-multiplicity jets with $\sigma_{p}=3$ mb and hFSI or with $\sigma_{p}=0$ mb and no hFSI.}
    \label{fig:v2VsDeta}
\end{figure}

Finally, we study the pseudorapidity-gap dependence of $v_2^*$. Increasing the pseudorapidity gap suppresses short-range nonflow correlations. As shown in Fig.~\ref{fig:v2VsDeta}, the calculations with final-state interactions decrease more slowly with increasing $|\Delta\eta^*_{\mathrm{min}}|$ than those without final-state interactions. This suggests that the anisotropy generated by final-state interactions is less sensitive to the pseudorapidity-gap selection.

\section{\label{sec:discussion} Discussion}
The present study supports a geometry--response picture for the emergence of collectivity inside high-multiplicity jets. As in the paradigm established for heavy-ion collisions, jets possess nonzero initial eccentricities, and final-state scattering converts the initial spatial anisotropy into the observed momentum anisotropy.

Within the present model setup, the anisotropy enhancement originates almost entirely from hadronic scattering, while the contribution from partonic scattering is negligible. This result should not be interpreted as evidence that partonic interactions are generally unimportant for collectivity inside jets. Rather, it reflects the limited partonic opacity in the current implementation of the transport model. In ZPC, partonic scattering begins only after the shower partons have completed their formation and is restricted to elastic $2\rightarrow2$ processes. As a result, the average partonic opacity remains small, even for the largest partonic cross section considered in this work. A more realistic description of jet evolution would allow parton branching and scattering to occur simultaneously throughout the shower evolution, potentially increasing the partonic opacity and modifying the relative contributions of the partonic and hadronic stages. Further developments in this direction will be important for quantitatively understanding the role of partonic scattering inside jets.

The jet-substructure dependence further highlights the interplay between the initial geometry and the strength of final-state interactions. Although the model predicts a larger anisotropy enhancement for one-prong jets than for two-prong jets, opposite to the current CMS measurement~\cite{CMS:PAS:HIN-24-024}, the enhancement increases with the initial eccentricity for both jet categories. At the same time, one-prong jets exhibit a systematically larger transport opacity. One possible explanation is that the two hard cores in two-prong jets are more widely separated in coordinate space, reducing the probability of rescatterings among their associated shower partons and leading to a smaller effective opacity. These observations indicate that the initial eccentricity and the transport opacity are both essential ingredients for the development of collectivity and may compete in determining the observed jet-substructure dependence.

Recently, Ref.~\cite{Dokshitzer:2025dr} proposed an alternative interpretation of the CMS measurement, attributing the observed enhancement of $v_2^*$ to a topology bias: high-multiplicity jets preferentially select two-prong jets, which possess a larger intrinsic azimuthal anisotropy while the two subjets remain causally disconnected throughout their evolution. Our results are partially consistent with this picture. In the absence of final-state interactions, two-prong jets exhibit a larger baseline $v_2^*$ than one-prong jets, confirming that the jet topology itself contributes to the observed anisotropy.
However, the topology bias alone does not account for the multiplicity-dependent enhancement observed by CMS. In the present transport model, the topology bias is already present without final-state interactions, yet the additional enhancement only emerges once rescatterings are included, indicating that interactions among the jet constituents play an essential role. Since high-multiplicity jets contain abundant soft particles produced through nonperturbative dynamics, a description based solely on perturbative shower evolution may miss important collective effects. Understanding the interplay between perturbative shower evolution and nonperturbative transport dynamics therefore remains an important open question.

\section{\label{sec:conclusion} Conclusion}
In this work, we investigated the microscopic origin of collective behavior inside high-multiplicity jets using a hybrid transport model that combines PYTHIA 8-generated jets with partonic and hadronic rescatterings. By following the full space--time evolution of the jet shower, we established a direct connection between the initial geometry, the subsequent transport evolution, and the final-state azimuthal anisotropy.

We found that high-multiplicity jets possess nonzero initial eccentricities that exhibit substantial longitudinal decorrelation. Nevertheless, the eccentricity correlation remains nonzero up to a pseudorapidity separation of at least $|\Delta\eta^*|=2$, suggesting that a Bjorken-like picture, in which particles separated in rapidity inherit a common initial geometry, may remain qualitatively applicable even in such a small system. The enhancement of the final-state anisotropy increases with the initial eccentricity, supporting a geometry--response mechanism inside jets.

Within the present transport model, the anisotropy enhancement is found to originate predominantly from hadronic rescatterings, while the contribution from partonic rescatterings remains negligible. This behavior is likely a consequence of the limited partonic opacity in the current implementation of ZPC, where rescatterings occur only after parton formation and are restricted to elastic $2\rightarrow2$ processes. Furthermore, although the predicted jet-substructure dependence differs from the current CMS measurement, our study suggests that the collective response is governed by both the initial geometry and the transport opacity. The weaker enhancement predicted for two-prong jets may indicate that the larger spatial separation between the two hard cores reduces the probability of rescatterings among their associated shower partons.

Overall, this study demonstrates that long-range correlations inside high-multiplicity jets can be understood within a geometry--response framework, where both the initial spatial anisotropy and the subsequent transport evolution play essential roles. At the same time, the remaining discrepancies with the CMS measurement highlight the need for more realistic descriptions of jet evolution, particularly the interplay between parton shower branching and transport dynamics. Such developments will provide deeper insight into the microscopic origin of collectivity in the smallest QCD systems.

\begin{acknowledgments}
This work was supported in part by the U.S. Department of Energy under Grant No. DE-SC0005131, the National Natural Science Foundation of China under Grant No. 12535010, and the U.S. National Science Foundation under Grant No. 2310021.
\end{acknowledgments}

\bibliography{apssamp}

\end{document}